\documentclass{amsart}
\usepackage{mathtools,amsthm,amssymb}

\usepackage{natbib}

\usepackage{xspace}
\usepackage{amsmath}
\usepackage{relsize}
\usepackage{nicefrac} 
\usepackage{enumitem}
\usepackage[pdftex]{graphicx}
\usepackage{subfig}
\usepackage{float}
\usepackage{hyperref}
\usepackage{tikz}

\usepackage{amsaddr}
\title[Data-driven CPE]{A data-driven change-point estimator}

\author{Stefanie Schwaar}
\address{Fraunhofer ITWM, Fraunhofer Platz 1, Kaiserslautern}
\email{stefanie.schwaar@itwm.fraunhofer.de}

\newtheorem{Theore}{Theorem}[section]
\newenvironment{Theorem}{\begin{Theore}}{\end{Theore}}

\newenvironment{Proof*}{\textbf{ Proof:}\\}{\hspace*{\fill}\rule{2mm}{2mm}\\[3mm]}

\newtheorem{Corollary}{Corollary}[section]

\newcommand{\argmax}{\operatorname*{arg\,max}}
\newcommand{\nconv}{\ensuremath{\underset{n\rightarrow \infty}{\longrightarrow}}}
\newcommand{\OP}{\ensuremath{O_{P}}}
\newcommand{\oP}{\ensuremath{o_{P}}}

\newcommand{\pconv}{\ensuremath{ \overset p {\longrightarrow}}}

\begin{document}
	
\thispagestyle{empty}

\pagenumbering{arabic}

\date{\today}

\begin{abstract} 
The q-weighted CUSUM and their corresponding estimator are well known statistics for change-point detection and estimation. They have the difficulty that the performance is highly dependent on the location of the change. An adaptive estimator with data-driven weights is presented to overcome this problem, and it is shown that the corresponding adaptive change-point tests are valid.
\end{abstract}

\maketitle


\section{Introduction}
Change-point analysis focuses on detection of structural breaks within the observations.
For {i.i.d.} observations with at most one change (AMOC) in mean (\citet[section 2.1]{bookCsorgoHorvath}, \citet{Antochetal}), in location (\citet[section 2.2]{bookCsorgoHorvath}, \citet{Huskova95}) or in the variance (\citet{GombayeA96}) test statistics and estimators were analysed and applied in many fields, from monitoring of industrial production or intensive-care patients to change detection in climate research or financial market states. More recently, those tests and estimators have been generalized to more complex models, e.g. to long-term dependent times series (\citet{Kraemer2001}, \citet{DehlinRoochTaqqu2013Nonpara}, \citet{HorvathKokoszka97}) or functional time series (overview is given in \citet{bookHorvathKokozka2012}). While change-point tests have been analysed for a wide range of models, the literature for change-point estimators is, however, comparatively scarce.
 \\
It is well-known that the quality of estimators and tests depend on the location of the changes where, in particular, change-points close to the boundaries of the observation period may be hard to detect resp. difficult to estimate. As a potential remedy for this problem, in the context of the common CUSUM test, corresponding to (\ref{eq:CPT}) below with $\gamma=0$, more general q-weighted CUSUM-type statistics have been proposed and analysed (for an overview see\citet{bookCsorgoHorvath}). This results in classes of corresponding estimators which perform well for various change-point locations. However, for choosing the right estimator, the unknown location would have to be known. In this paper, we propose and investigate a data-driven method for choosing the weights. We show that the estimator resulting from maximizing the adaptively weighted CUSM statistic is consistent and has the right rate of convergence (Corollary \ref{Cor:ResultEstimator}). We illustrate with some simulations that this data-driven estimate shows a good performance independently from the location of the change-point in contrast to the standard estimator using fixed weights.
 \\
We expect the adaptively weighted CUSUM tests to also show a more uniform behaviour regarding power. Analysing the performance of tests and estimators under the alternative \citet{Vogelsang99} investigated causes for non-monotonic power. The nuisance variance is one possible reason and self-normalized change-point tests are analysed by \citet{ZhangShao2010} and \citet{ZangLavitas2018}. Those statistics have pivotal limit distribution under the no change hypothesis and monotone power under the alternative. This means that the power increases for increasing magnitude. Considering a fixed magnitude of the change, the power, however, differs for different change-point locations and may be quite bad, e.g. for the common CUSUM test if the change happens early or late. To the best knowledge of the author, a change-point test and estimator performing uniformly (w.r.t. to the location of the change) best is not given so far. Our long-term goal is to derive a test statistic for which the empirical power is independent on the location of the change. As a first step, we also show in this paper that the adaptively weighted CUSUM test attains asymptotically the prescribed level (Corollary \ref{Cor:FirstReasultofAsymptotic_Teststatistik}) and is consistent under the alternative (Corollary \ref{Cor:secondReasultofAsymptotic_Teststatistik}) such that it is a valid alternative to the common test with fixed weights. \\
The next section gives a brief description of the test statistics and estimators considered here. Section \ref{plugin} focuses on the plug-in estimator and its properties, and some theory for the corresponding change-point estimators and tests are also considered. Section \ref{sec:sim} contains a simulation study showing that the data-driven weighted change-point estimator exhibits the desired behaviour.

\section{The weighted CUSUM statistic}
To illustrate our approach, we focus on the at most one change in mean model with {i.i.d.} observations, i.e.  
\begin{equation}\label{eq:model}
X_i=\begin{cases}
\mu +\varepsilon_i & ,\ 1\le i\le m\,,\\
\mu+\delta_n+\varepsilon_i &,\ m < i\le n\,,
\end{cases}
\end{equation}
where $m$ denotes the unknown change-point, $\delta_n$ the unknown size of the change and  $\varepsilon_t$ are the centred residuals with unknown finite second moment. The change-point  $m$ is given as   $m=\lfloor \tau n\rfloor$ with $\tau \in (0,1]$, i.e. for $\tau=1$ there is no change ($H_0$) and for $\tau\in (0,1)$ there is a change ($H_1$). The size of the change, $\delta_n$, is assumed to fulfil that $\delta_n^2\neq 0$ is non-increasing and $n\delta_n^2\nconv \infty$. Following \citet{bookCsorgoHorvath} who gave a detailed analysis of this well known model, we consider the CUSUM-type statistic 
\begin{equation}\label{eq:CPT} T_n(\gamma)=\max_{1\le k<n}w_{\gamma}(\mathsmaller{{k}/{n}}) S_n(k), \quad \text{where } \; \; S_n(k) = \frac{1}{\sqrt{n}}
	\left|\sum^k_{i=1}(X_i-{\hat \theta})\right|\,
	\end{equation}
	with $w_\gamma(s)=(s(1-s))^{-\gamma}$, $\gamma\in[0,1/2]$,
and the corresponding change-point estimator 
\begin{equation}\label{eq:CPE}
{\hat m}_{\gamma}=\argmax_{1\le k<n}w_{\gamma}(\mathsmaller{{k}/{n}}) S_n(k) \,.
\end{equation}
Based on well-known results, we derive a plug-in estimator $\hat{\gamma}$ for $\gamma$ and show asymptotic results for the change-point test and estimator.\\
Observe, the rescaled change-point estimator $\hat{\tau}={\hat m}_{\gamma}/n$ only depends on the estimate ${\hat m}_{\gamma}$. Point estimates ${\hat m}_{\gamma}$ have been studied for a variety of complex change-point problems and regularity conditions are known (see \citet{KirchKamgaing2015} or \citet{DisSchwaar}).\\

\section{Plug-in technique} \label{plugin}
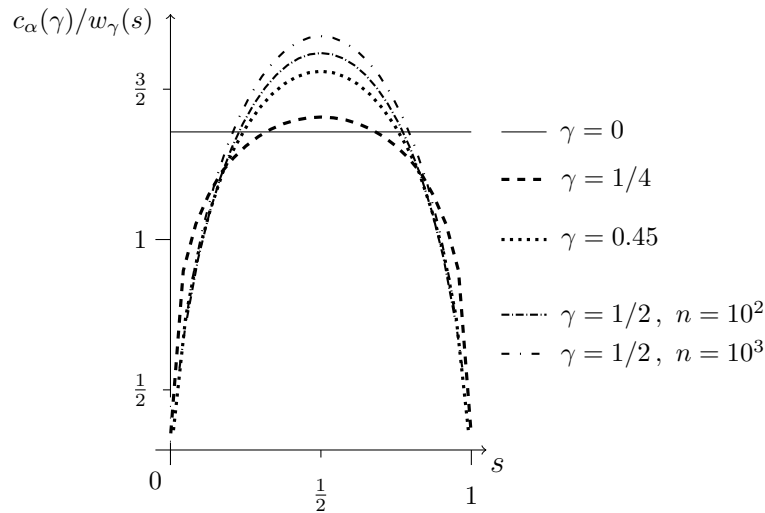
\begin{figure}[h]
\begin{center}
\begin{tikzpicture}[liniendicke/.style={#1},scale=4]
\draw[->] (-0.05,0.3) -- (1.05,0.3) coordinate (x axis);
\draw[->] (0,0.475) -- (0,1.75) coordinate (y axis);
\draw[loosely dotted] (0,0.25)--(0,0.475);
\draw (0,0.25) -- (0,0.325);

\draw (-0.05,0.3-0.1) node {0};
\draw (1.08,0.3-0.05) node {$\, {\mathlarger {\mathlarger s}}$};
\draw (-0.025,1.725) node[left] {${ {\mathlarger c_\alpha(\gamma)}/{\mathlarger w_{\gamma}(s)}}$};

\draw (1,0.3)--(1,0.3-0.05);
\draw (1,0.3-0.15) node {1};

\draw (0.5,0.3)--(0.5,0.3-0.025);
\draw (0.5,0.3-0.15) node {$\frac{1}{2}$};

\draw (0,0.5)--(-0.025,0.5);
\draw (-0.1,0.5) node {$\frac{1}{2}$};
\draw (0,1)--(-0.05,1);
\draw (-0.1,1) node {1};
\draw (0,1.5)--(-0.025,1.5);
\draw (-0.1,1.5) node {$\frac{3}{2}$};

\draw [densely dash dot,liniendicke/.expanded=thick,domain=0.055:0.945] plot (\x,{sqrt(\x*(1-\x))*1/0.3085634}) ;
\draw [densely dash dot,liniendicke/.expanded=thick,](1.1,0.75) --(1.25,0.75) node[right] {$\,\gamma={1}/{2}\,,\ { n=10^2}$};
\draw [loosely dash dot,liniendicke/.expanded=thick,domain=0.03:0.97] plot (\x,{sqrt(\x*(1-\x))*1/0.2982199}) ;
\draw [loosely dash dot, liniendicke/.expanded=thick,domain=1.1:1.25] plot (\x,0.62) node[right] {$\,\gamma={1}/{2}\,,\ { n=10^3}$};

\draw [dashed,liniendicke/.expanded=very thick, domain=0.001:0.999] plot (\x,{((((\x*(1-\x)))^(0.25))*(1.99)});
\draw [dashed,liniendicke/.expanded=very thick] (1.1,1.2) -- (1.25,1.2) node[right] {$\,\gamma={1}/{4}$};
\draw [dotted,liniendicke/.expanded=very thick,domain=0.01:0.99] plot (\x,{((\x*(1-\x))^(0.45))*2.91});
\draw [dotted,liniendicke/.expanded=very thick,domain=1.1:1.25] plot (\x,1)node[right] {$\,\gamma=0.45$};
\draw [domain=0:1] plot (\x,{(1.358)});
\draw (1.1,1.358) -- (1.25,1.358) node[right] {$\,\gamma=0$};

\end{tikzpicture}
\end{center}
\caption{
	This Graphic shows the critical function for $\alpha=0.95$ ($c_\alpha=1.358$ for $\gamma=0$, for $\gamma\in (0,1/2)$ the $c_\alpha$ are simulated by the adaptive Monte Carlo method).}
\label{fig:weightfunctions}
\end{figure}

The performance, i.e. the power of the test based on statistic \eqref{eq:CPT} or the mean-squared error of the corresponding estimator, depends on the location of the change. To illustrate this effect, Figure \ref{fig:weightfunctions} shows the critical line which the standard CUSUM statistic $S_n(k)$ has to cross for detecting a change-point. This shows, for early changes a test statistic with $\gamma$ close to $0.5$ is more sensitive. On the other hand, the standard CUSUM test ($\gamma=0$) is particularly sensitive in the case of a change in the middle. 

\begin{figure}[h]
	\begin{center}
		\includegraphics[scale=0.5]{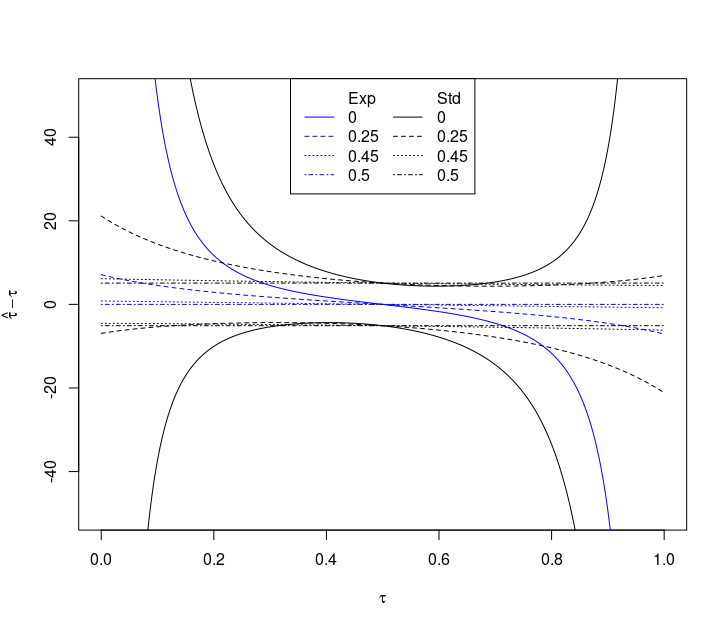}
	\end{center}
	\caption{ The graphic shows the expectation (blue) and standard deviations (black) of the asymptotic distribution for the w-weighted CUSUM statistic based on \citep{STRYHN1996}.}
	\label{fig:Exp_Std_Stryhn}
\end{figure}

Based on the asymptotic distribution of the change-point estimator the weighted CUSUM is preferable (compare Figure \ref{fig:Exp_Std_Stryhn}). While this is the asymptotic behaviour for finite observations the change-point estimator behaves analogously to the change-point test. A well known criteria for comparing estimators is the mean squared error. Here a comparison of the finite distribution, using the estimated density, and the MSE of the change-point estimators gives much more insides (see appendix \ref{Appendix:Esti_Cusum_w_Cusum}). We observe that for a change in the middle the estimator based on the standard CUSUM statistic is preferable to the weighted CUSUM. For early (or late) changes it is the other way around.  \\

We are interested in deriving an estimator and in the long run a test performing independently of the location. Additionally the test statistic should have an asymptotic distribution allowing for analytic calculation of quantiles. To achieve this goal, we use a plug-in technique for replacing $\gamma$ with an estimator.

\subsection{Weight Estimator}
Note, under $H_0$ an analytic formula for quantiles of the asymptotic distribution is known if $\gamma=0$. Under $H_1$ the plug-in estimator should be close to $0.5$ in case of early or late changes, and $0$ in the case of a change in the middle. Since, the asymptotic distribution of the weighted CUSUM based change-point estimator under $H_0$ is
\begin{equation}\label{eq:weightedCUSUM_CPE_H0}
\hat{\tau}:=\frac{{\hat m}_{\frac{1}{2}}}{n}\overset{d}{\longrightarrow} \xi
\end{equation}
with $P(\xi=0)=P(\xi=1)=\frac{1}{2}$ (Theorem 1.6.1 in \citet{bookCsorgoHorvath}), we use this estimator 
 for deriving the plug-in estimator. As we know, under $H_1$ (i.e. $\tau\in(0,1)$) it holds
\begin{equation}\label{eq:weightedCUSUM_CPE_H1}
\hat{\tau}-\tau=\oP(1)\,,
\end{equation}
the following estimator is used
\begin{equation}\label{eq:hat_gamma}
\hat{\gamma}=g(\hat{\tau})\,,
\end{equation}
with a function $g:[0,1]\rightarrow[0,0.5]$ fulfilling suitable regularity conditions.\\

\begin{Theorem}\label{theo:Replacement}
	For $g:[0,1]\rightarrow[0,0.5]$ a Lipschitz-continuous function and $\hat{\tau}$ as in \eqref{eq:weightedCUSUM_CPE_H0}, i.e. $\hat{\tau}\in[0,1]$. Then
	\begin{enumerate}[label=\alph*),ref=\theTheore.\alph*)]
		\item \label{theo:Replacement_w_HO}under $H_0$ and for $g(0)=g(1)=0$, we have
	$$ \max_{1\le k < n} |w_{g(\hat{\tau})}(k/n)-1| =\oP(1)\,.$$
	\item \label{theo:Replacement_w_HI}under $H_1$ and for $\delta_n^{-2}n^{-1}\log(n)\nconv 0$, we have
	$$ \max_{1\le k < n} \left|\frac{w_{g(\hat{\tau})}(k/n)}{w_{g(\tau)}(k/n)}-1\right| =\oP(1)\,.$$
\end{enumerate}
\end{Theorem}

\begin{Proof*}
	\begin{enumerate}[label=\alph*)]
		\item From $w_{\gamma}(k/n)\ge 1$ for all $k=1,\dots,n-1$ and $\gamma\in[0,\infty)$, it follows
		\begin{eqnarray} \label{eq:proof_H0_weight}
		\max_{1\le k < n} |w_{g(\hat{\tau})}(k/n)-1| &=& \max_{1\le k < n} (w_{g(\hat{\tau})}(k/n))-1 \nonumber\\
		&=&\left(\frac{1}{n}\left(1-\frac{1}{n}\right)\right)^{-g(\hat{\tau})}-1\,.
		\end{eqnarray}
		If and only if 
		$g(\hat{\tau})(\log(n)+\log(\frac{n}{n-1}))=\oP(1)$, we have $
		(\frac{1}{n}(1-\frac{1}{n}))^{-g(\hat{\tau})}-1=\oP(1)$.
		From \eqref{eq:weightedCUSUM_CPE_H0} and the Lipschitz-property, it follows
\begin{align*}
P\Bigg({g}\left(\hat{\tau}\right)\log(n)<\epsilon\Bigg)
\ge 
&P\left(\hat{\tau}<\frac{\epsilon^\prime}{\log(n)},\hat{\tau}<\frac{1}{2}\right)\\
&+P\left(1-\hat{\tau}<\frac{\epsilon^\prime}{\log(n)},1-\hat{\tau}<\frac{1}{2}\right)\,
\end{align*}
with $\epsilon^\prime=\epsilon/C$ and $C$ the Lipschitz constant. 
Analog to the proof of \cite[Theorem 1.6.1]{bookCsorgoHorvath} we have for $n$ large enough,
$$P\left(\hat{\tau}<\frac{\epsilon}{\log(n)}\right)+P\left(1-\hat{\tau}<\frac{\epsilon}{\log(n)}\right)\nconv 1\,.$$
This yields
$$P({g}(\hat{\tau})\log(n)<\epsilon)\nconv 1$$
for all $\epsilon>0$.
		\item 	We know from \citet{Antochetal} that $\hat{\tau}-\tau=\OP(\delta_n^{-2}n^{-1})$, if 
		$$ \delta_n\nconv 0,\qquad \delta_n^{-1}n^{-\frac{1}{2}}(\log\log n)^{\frac{1}{2}}\nconv 0\,.$$
		It follows that $g(\hat{\tau})-g(\tau)=\OP(\delta_n^{-2}n^{-1})$ and so
		$$ |g({\tau})-g(\hat{\tau})|(\log(n)+\log(\frac{n}{n-1}))=\oP(1)\,.$$
		This implies directly 
		$$\left|\left(\frac{1}{n}\left(1-\frac{1}{n}\right)\right)^{g({\tau})-g(\hat{\tau})}-1\right|=\oP(1)\,,$$
		and the assertion follows from an argument analogously to \eqref{eq:proof_H0_weight}.
	\end{enumerate}
\end{Proof*}

\subsection{Change-point technique}
We consider the $w$-weighted CUSUM-type \linebreak statistic \eqref{eq:CPT}
 and the corresponding change-point estimator \eqref{eq:CPE}.
 In \citet{KirchKamgaing2015} and \cite[section 3.4 and 4.4]{DisSchwaar} regularity conditions are given for this type of change-point test and estimator.

Note that 
\begin{align*}\label{eq:main_Idea_CPA}
w_{g(\hat\tau)}(k/n)\frac{1}{\sqrt{n}}\left|\sum^k_{i=1}(X_i-{\hat \theta})\right|
&=(1+\oP(1))w_{g(\tau)}(k/n)\frac{1}{\sqrt{n}}\left|\sum^k_{i=1}(X_i-{\hat \theta})\right|\,.
\end{align*}
directly implies the following Corollary by standard arguments based on Slutsky's Lemma.\\
Consider the \textbf{change-point test} \eqref{eq:CPT}, then Theorem \ref{theo:Replacement_w_HO} yields the following result.
\begin{Corollary}
	\begin{enumerate}[label=\alph*),ref=Corollary\ \theCorollary\ \alph*)]
		\item \label{Cor:FirstReasultofAsymptotic_Teststatistik}	Under $H_0$ and $g:[0,1]\rightarrow [0,0.5]$ being Lipschitz continuous with $g(0)=g(1)=0$, we have 
		\begin{equation*}
		T_n(g(\hat{\tau}))-T_n(0)=\oP(1)\,.
		\end{equation*}
		\item \label{Cor:secondReasultofAsymptotic_Teststatistik}	Under $H_1$, $g:[0,1]\rightarrow [0,0.5]$ Lipschitz continuous and $\delta_n^{-2}n^{-1}\log(n)\nconv 0$, we have
		\begin{equation*}
		T_n(g(\hat{\tau}))\pconv \infty\,.
		\end{equation*}
	\end{enumerate}
\end{Corollary}
Observe, in \ref{Cor:FirstReasultofAsymptotic_Teststatistik} the quantile of the Kolmogorov distribution is given and therefore known.\\

	An other approach would be to use the plug-in estimator also for the quantile, instead of making use of the well known asymptotic distribution.	Equivalent to \ref{Cor:FirstReasultofAsymptotic_Teststatistik}, it follows $|q_{1-\alpha}(g(\hat{\tau}))-q_{1-\alpha}(g(\gamma))|=\oP(1)\,,$
	where $	q_{1-\alpha}(g(\gamma))$ denotes the $1-\alpha$ quantile of the asymptotic distribution of the test statistic with weight function $w_{g(\gamma)}$.
 In application, these quantiles need to be calculated with Monte-Carlo simulation. The standard Monte-Carlo simulation of the quantiles for weight function with $g(\hat{\tau})$ close to $0.5$ takes is not pratical as it takes way to much time. Therefore, a current research topic is finding a applicable simulation method for those weight functions.\\

For the \textbf{change-point estimator} \eqref{eq:CPE}, we conclude analogously.
\begin{Corollary} \label{Cor:ResultEstimator}
	Under the conditions of Theorem \ref{theo:Replacement_w_HI}, it follows that the estimator having a data-driven weight function is still a consistent estimator and the convergence rate does not change compared to a fixed weight function.
\end{Corollary}
Notize, it is clear that the results hold true for all $w_\gamma$ weighted CUSUM-type statistics where the corresponding $w_{1/2}$-weighted change-point estimator fulfils \eqref{eq:weightedCUSUM_CPE_H0}, \eqref{eq:weightedCUSUM_CPE_H1} and $P({\hat m}_{1/2}/n<\epsilon/\log(n))\nconv 1$ for all $\epsilon>0$.

\section{Simulation study}\label{sec:sim}

\subsection{Change-point estimator}
\begin{figure}[H]
	\begin{center}
		\includegraphics[scale=0.35]{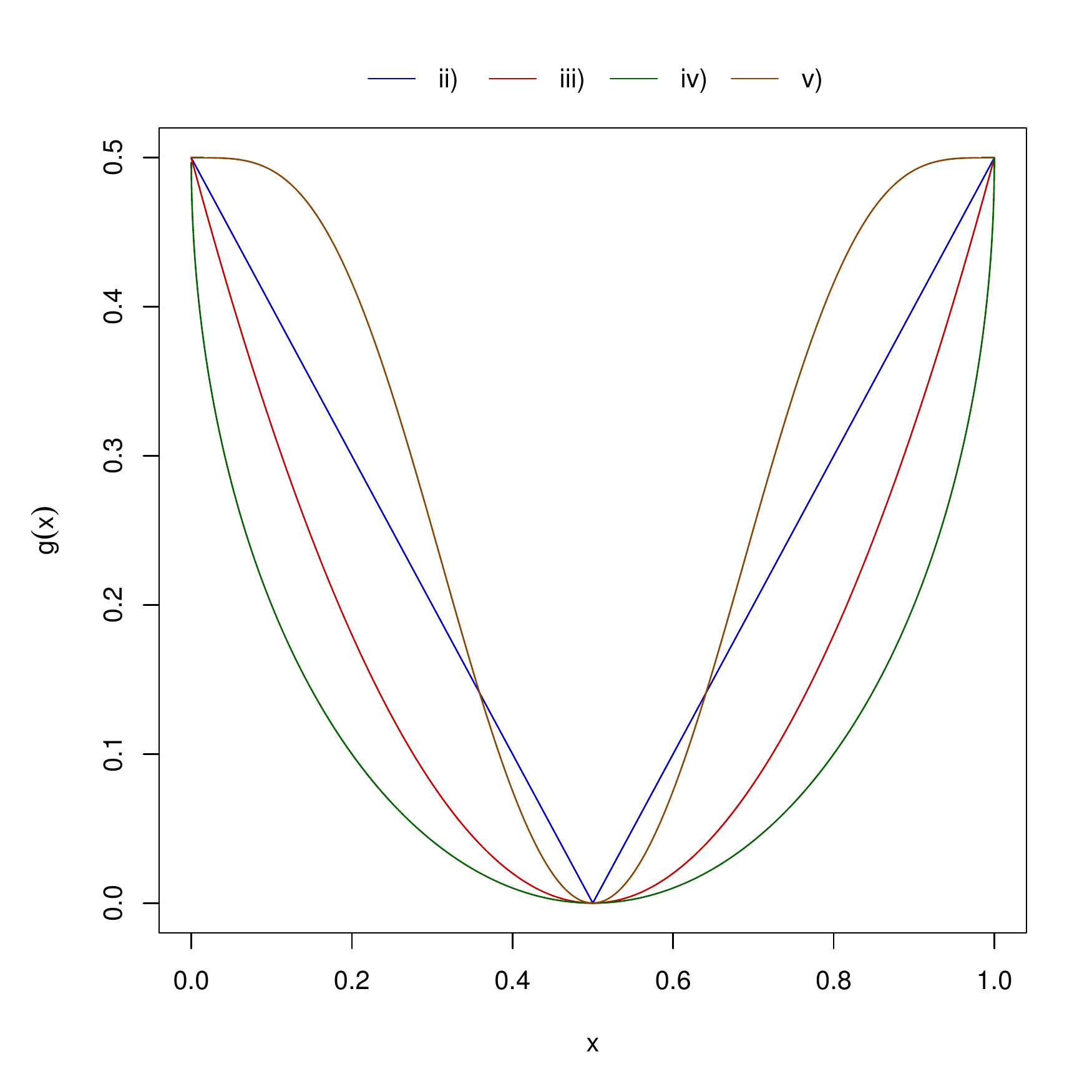}
	\end{center}
	\caption{Functions used for replacing}
	\label{fig:function_g_CPE}
\end{figure}

For the simulation study, we choose four different functions $g$ fulfilling the regularity conditions.  We consider 4 different types of continuous functions $g$ on $[0,0.5]$, fulfilling \begin{equation}\label{eq:Conditions_gamma_H1}
\begin{split}
\text{early/late changes:}\qquad &\gamma\approx 0.5\,,\\
\text{change in the middle}\qquad & \gamma=0\,.
\end{split}
\end{equation}
As Benchmark we include the cases $g(x)\equiv0$ and $g(x)\equiv 1/2$ and therefore consider the following change-point estimator:
\begin{align*}
i)& \ g(x)=0, &
{ii)}&\ g(x)={|x-0.5|},\\
{iii)} &\ g(x)={0.5-}2x(1-x), & 
{iv)} &\ g(x)={0.5-}\sqrt{x(1-x)},\\
{v)}&\ g(x)={0.5-}2^7(x(1-x))^4 &
vi),&\ g(x))=\frac{1}{2}.
\end{align*}
Figure \ref{fig:function_g_CPE} shows those functions. 

To analyse the performance of the change-point estimator we consider the mean squared error of the rescaled change-point estimator and the density function. We consider four different distributions in the simulation study: standard normal ($\mathcal{N}(0,1)$), exponential ($Exp(1)$), poisson ($Poi(1)$) and uniform ($U[0,1]$). Here we present the results for the standard normal distribution. The corresponding results for the other distributions are comparable and given in the appendix \ref{Appendix:Esti_data-driven}. The size of the underlying change is chosen within the range of $\delta\in[0.3,1.1]$. We get the following results.\\

\begin{figure}[h]
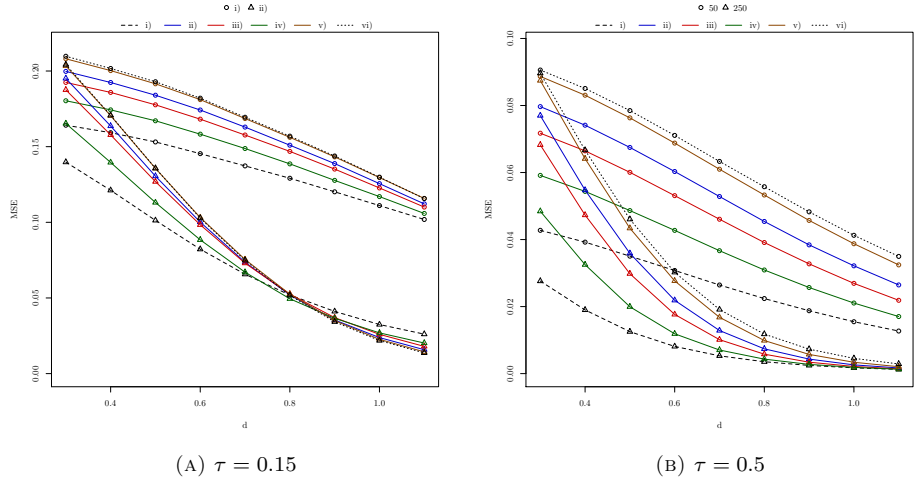

	\subfloat[$\tau=0.15$]{
		\resizebox{0.48\textwidth}{!}{\input{stdNormal_MSE_lambda=15pc.tex}}
	}
	\subfloat[$\tau=0.5$]{
		\resizebox{0.48\textwidth}{!}{\input{stdNormal_MSE_lambda=50pc.tex}}
	}
	\caption{empirical MSE ($M=10^5$ repetitions) of the plug-in change-point estimator for i.i. $\mathcal{N}(0,1)$-distributed observations with change of size $d$ at $\tau=0.15$ (A) or $\tau=0.5$ (B) }
	\label{fig:MSE_CPE_earlyChange_standardNormal}
\end{figure}

Figure \ref{fig:MSE_CPE_earlyChange_standardNormal} shows the empirical mean squared error of the change-point estimator for an early change. We used $M=10^5$ samples of length $N=50$ with change at $\tau=0.15$ and $\tau=0.5$.\\
\begin{figure}[h]
	\subfloat[$N=50$, $\tau=0.15$]{
		\includegraphics[scale=0.35]{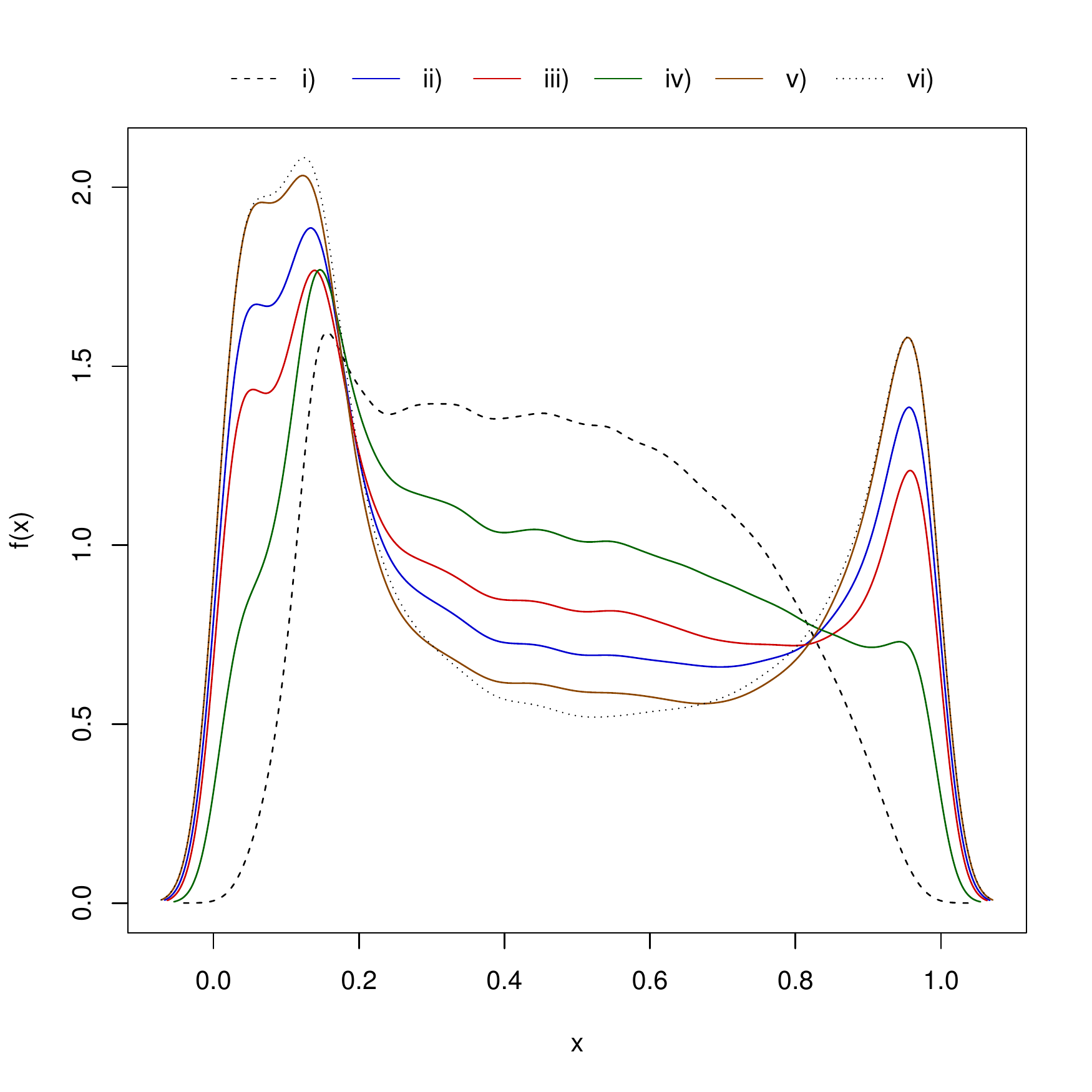}
	}
	\subfloat[$N=50$, $\tau=0.5$]{
		\includegraphics[scale=0.35]{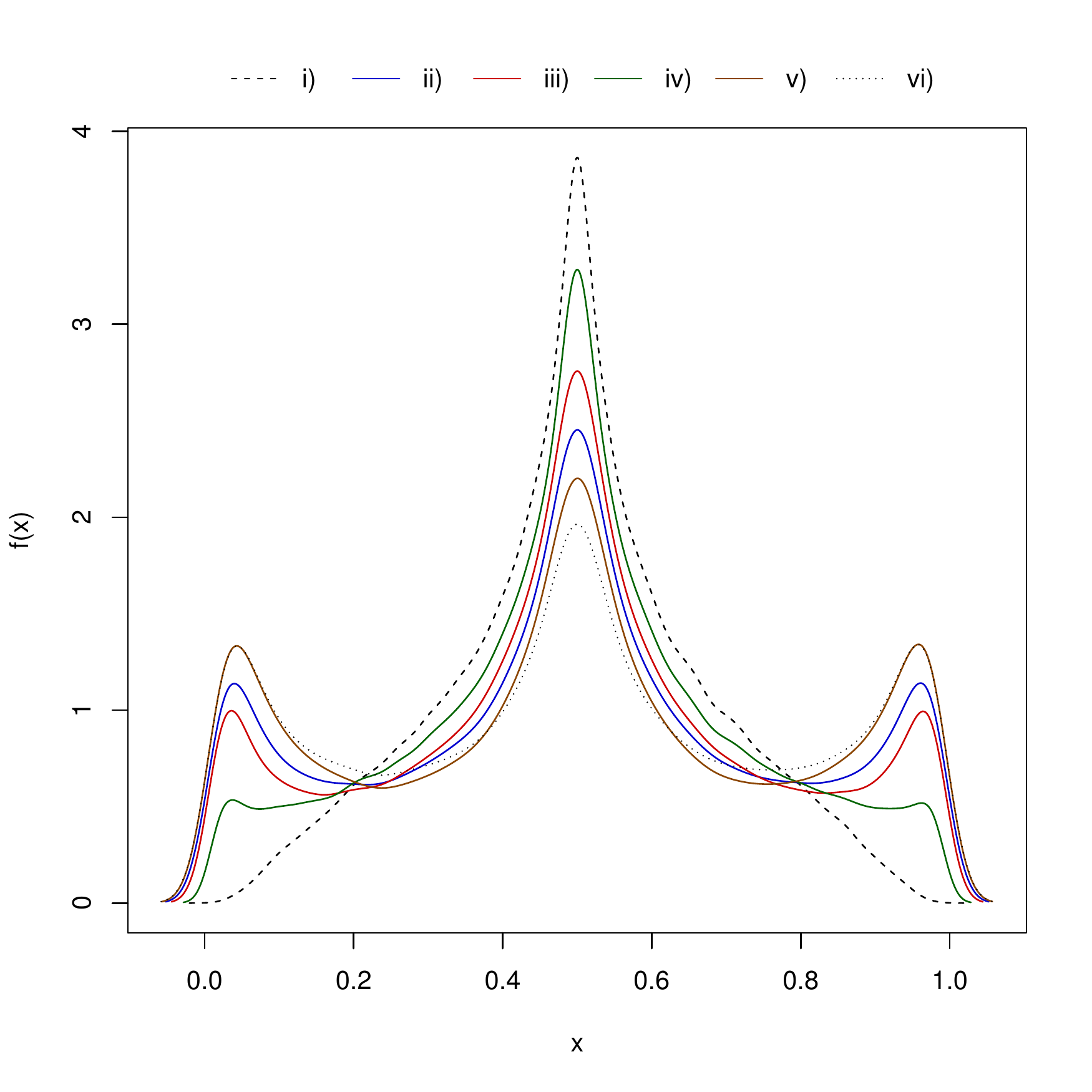}
	}\\
	\subfloat[$N=250$, $\tau=0.15$]{
		\includegraphics[scale=0.35]{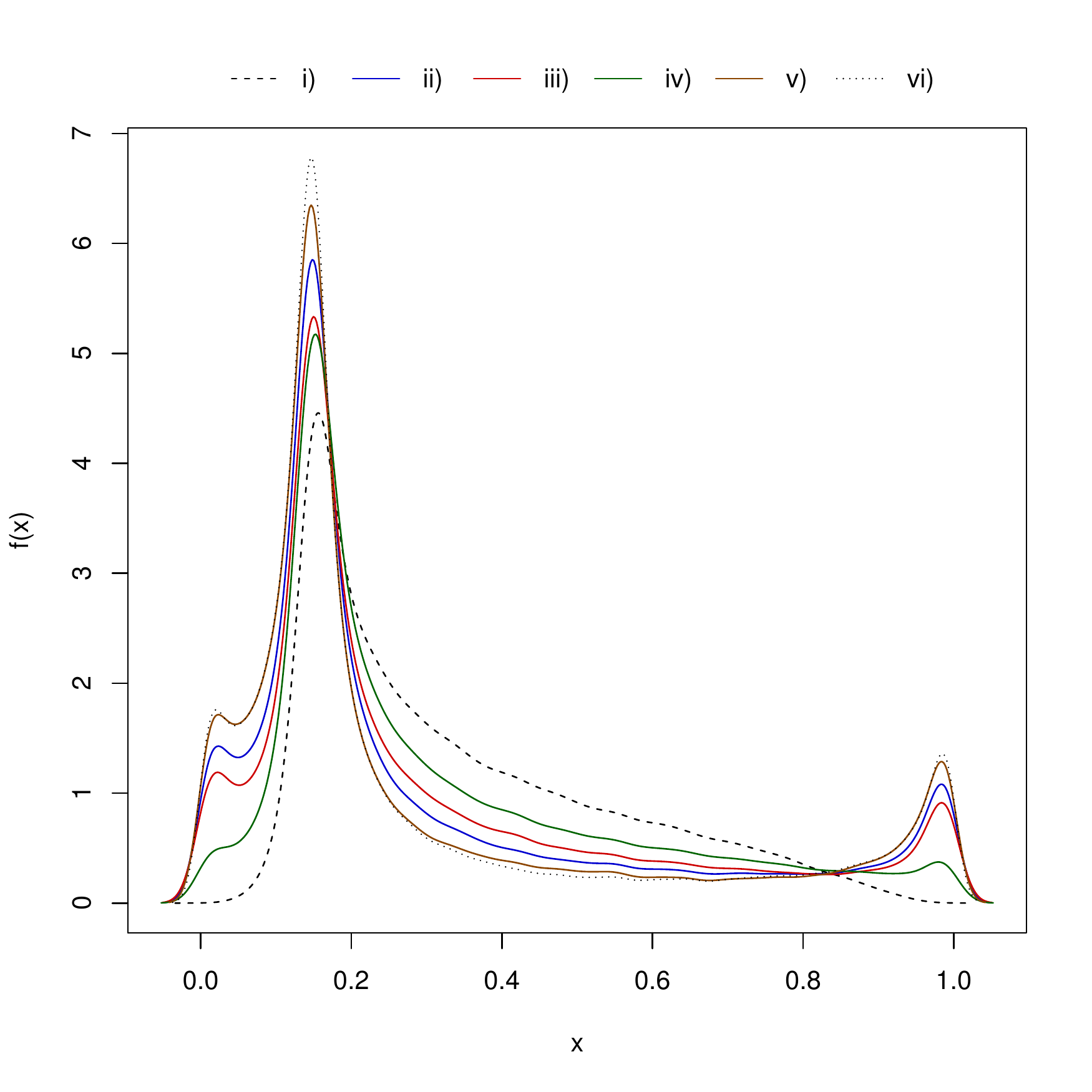}
	}
	\subfloat[$N=250$, $\tau=0.5$]{
		\includegraphics[scale=0.35]{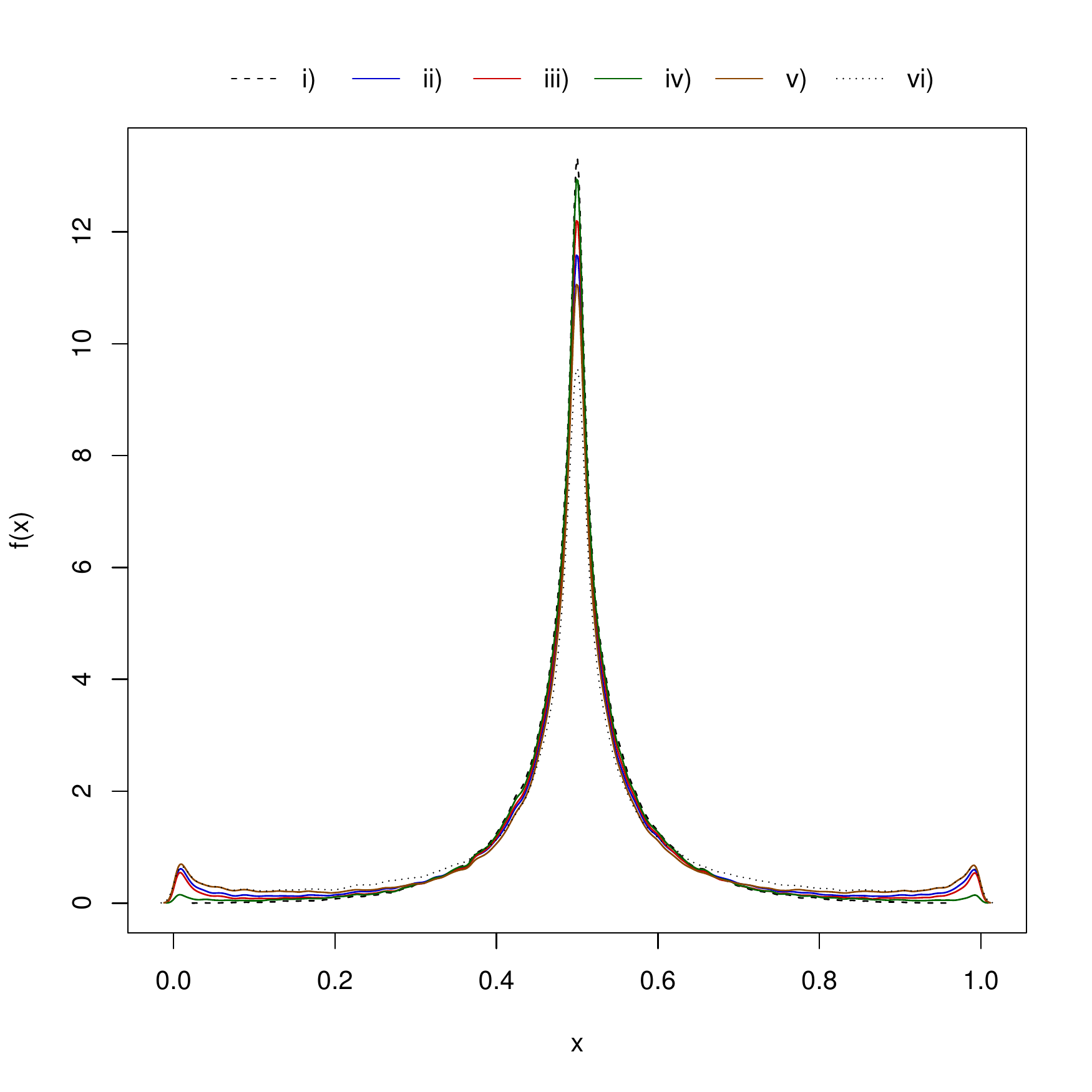}
	}
	\caption{empirical MSE ($M=10^5$ repetitions) of the plug-in change-point estimator for i.i. $\mathcal{N}(0,1)$-distributed observations with change of size $d=0.4$ at $\tau=0.15$ (A) and (C) or $\tau=0.5$ (B) and (D) }
\end{figure}

In the case of an early change, the estimator based on the standard CUSUM statistic ($i)$\,) overestimates the time point especially for small changes, while the estimator based on the usual weighted CUSUM ($vi)$) gives mostely good estimations. We also observe, in some cases the weighted CUSUM ($vi)$) gives estimation on the contrary position of the interval $[0,1]$, like the true value is at $\tau=0.1$ then in some cases we get  the estimations arround $0.9$. The proposed data-driven weighted change-point estimator instead finds a balance between the behaviour of those both statistics. While for the early change (Figure \ref{fig:MSE_CPE_earlyChange_standardNormal}) the functions $iv)$ or $iii)$ are preferable to $v)$, in the case of a change in the middle $v)$ would give better results than $iv)$ and $iii)$.\\

The discussed results are representative for the behaviour of the plug-in change-point estimator. For further simulation results  see Appendix \ref{Appendix:Esti_data-driven}. In conclusion, the performance is better for the data-driven change-point estimator using $g(x)=0.5-\sqrt{x(1-x)}$. The question whether this is the optimal choice of the function $g$ is still open for future research.

\newpage
\section{Conclusion and discussion}

	This publication deals with $q$-weighted CUSUM statistic and proposes a data-driven weighted CUSUM statistic. Therefore, we provide theoretical results for the data-driven weight function using the plug-in method.
For Change-point estimator we considered different functions for the data-driven weight. Those balance between the CUSUM and weighted CUSUM. The simulation study shows advantage for small sample sizes and/or small size of change. We proposed a data-driven weight function which shows the best balance. Still open is the proof of an uniformly best estimator and and is a current research topic. 

\section*{Acknowledgements}
At first the author wish to thank Prof. Dr. J\"urgen Franke and Prof. Dr. Ralf Korn for their suggestions. The author also wish to thank the DFG for funding within the RTG 1932 "Stochastic Models for Innovations in the Engineering Sciences".

\bibliography{citation}
\bibliographystyle{plainnat} 

\newpage
\begin{appendix}
\section{Cusum vs weighted Cusum}\label{Appendix:Esti_Cusum_w_Cusum}
~\\
\begin{figure}[H]
	\subfloat[$N=50$, $\tau=0.15$]{
		\includegraphics[scale=0.35]{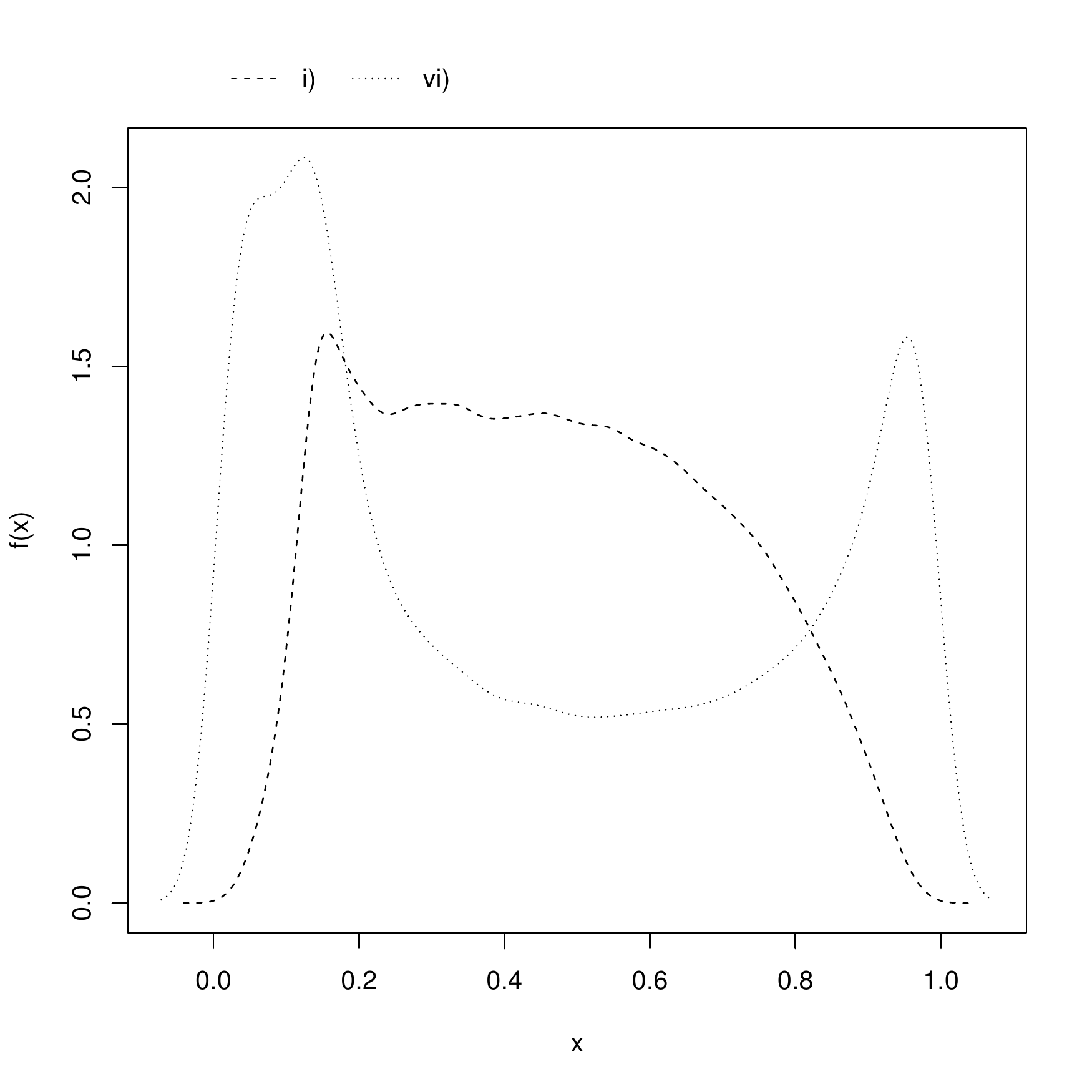}
	}
	\subfloat[$N=50$, $\tau=0.5$]{
		\includegraphics[scale=0.35]{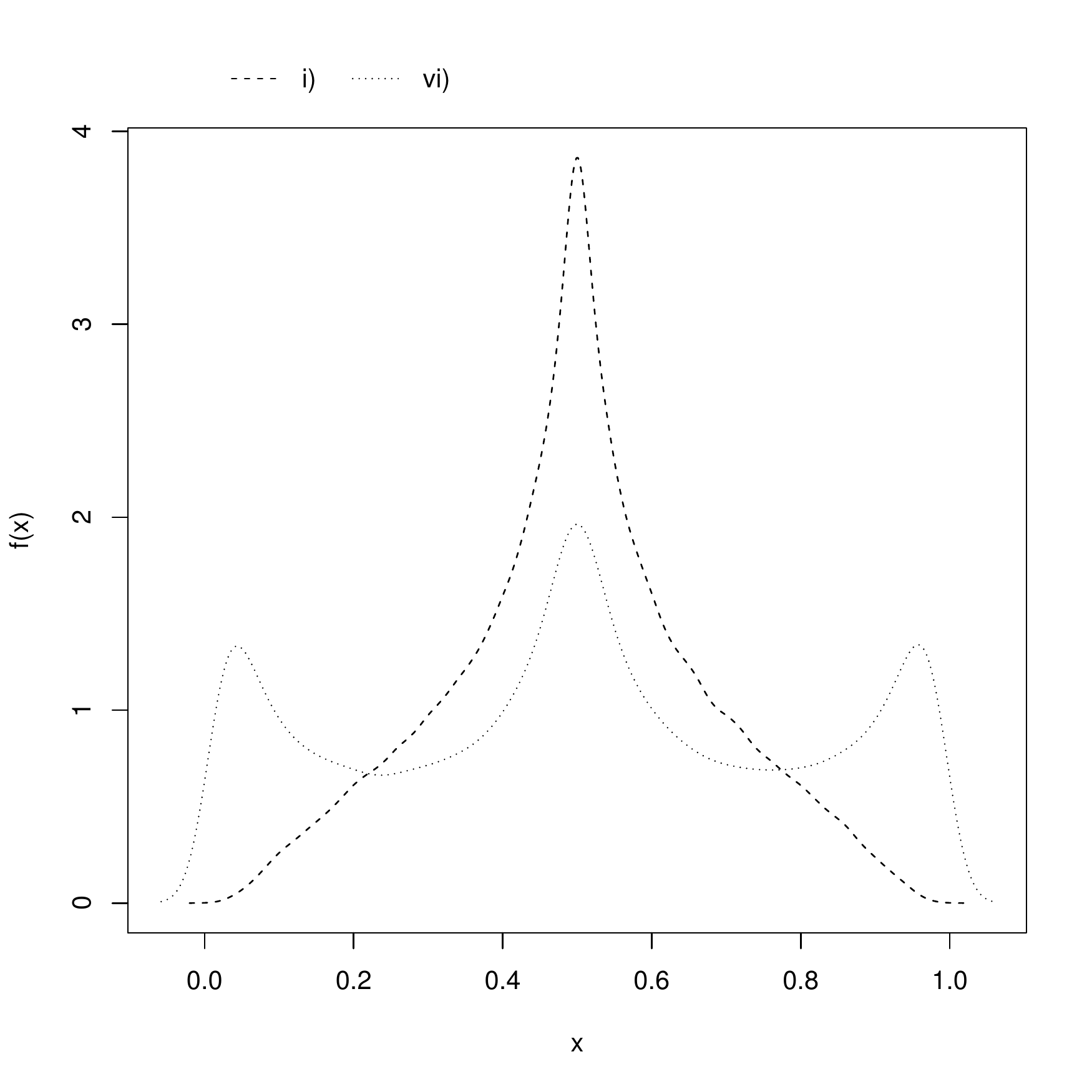}
	}\\
	\subfloat[$N=250$, $\tau=0.15$]{
		\includegraphics[scale=0.35]{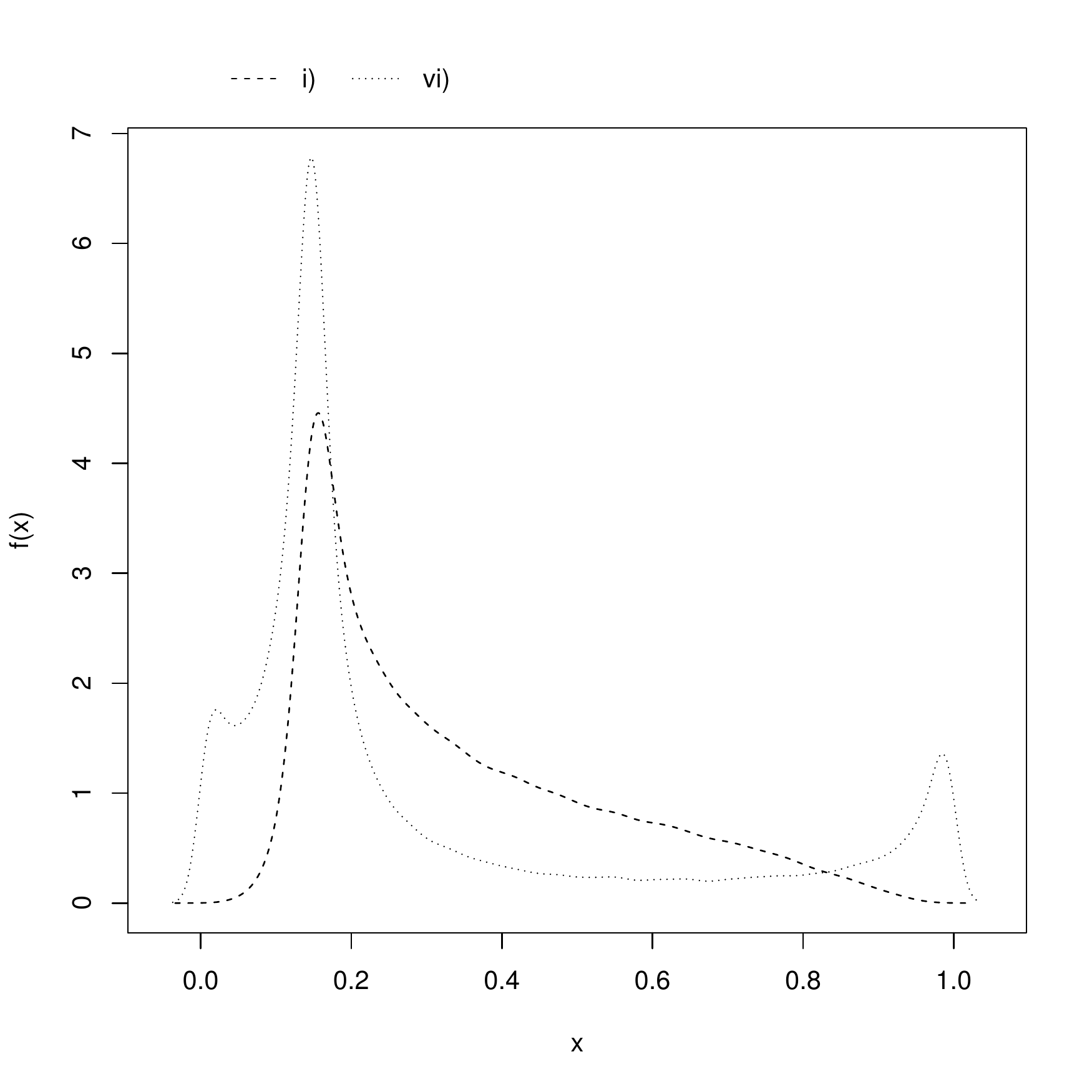}
	}
	\subfloat[$N=250$, $\tau=0.5$]{
		\includegraphics[scale=0.35]{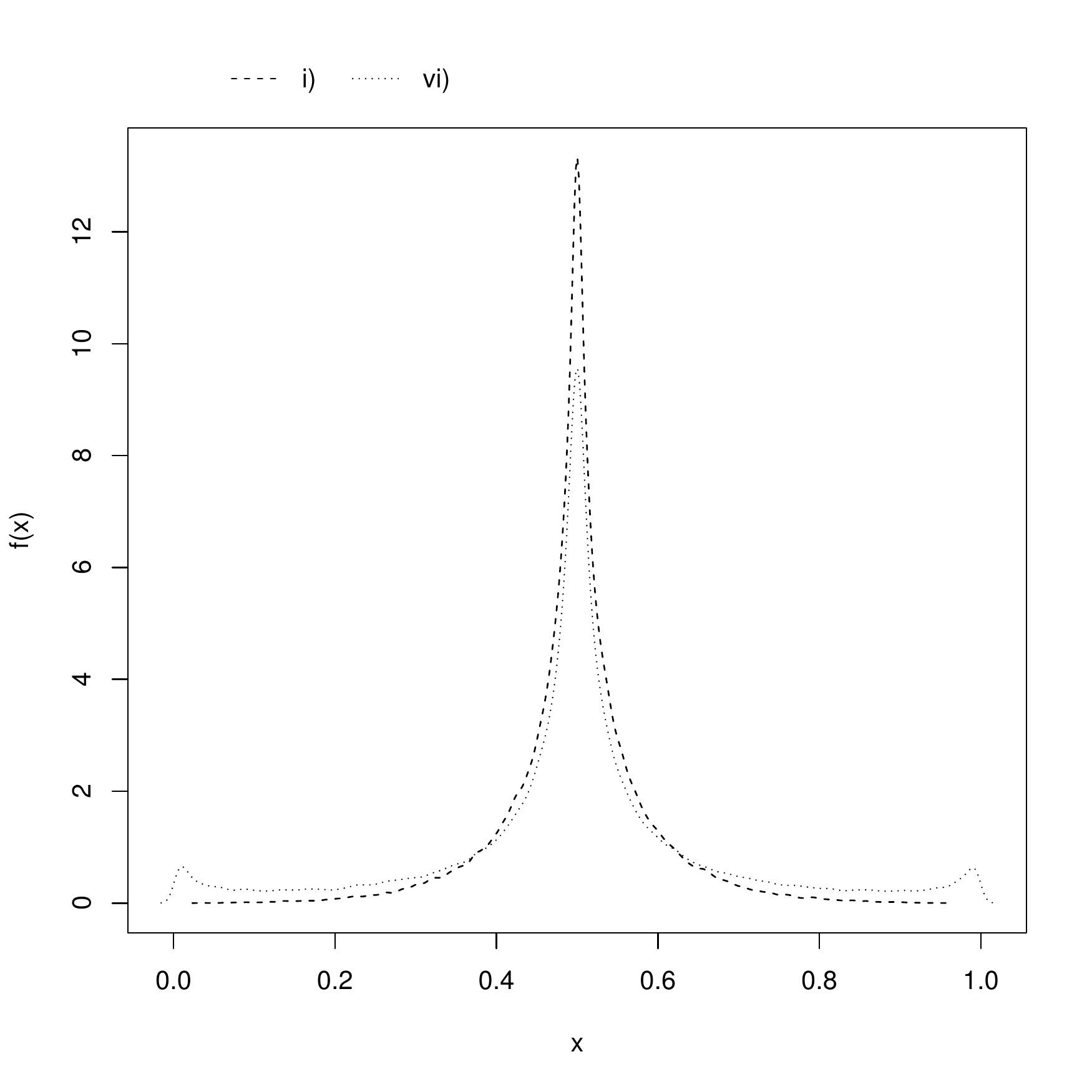}
	}
	
	\caption{rescaled density of Cusum and weighted Cusum change-point estimator based on $M=10^5$ simulations of $N$ i.i. $\mathcal{N}(0,1)$-distributed random variables with change of size $d=0.4$ at $\tau$}
	\label{fig:Density_CPE_Finite}
\end{figure}

\begin{figure}[H]
	\subfloat[ $\tau=0.15$]{
		\resizebox{0.48\textwidth}{!}{
\begin{tikzpicture}[x=1pt,y=1pt]
\definecolor{fillColor}{RGB}{255,255,255}
\path[use as bounding box,fill=fillColor,fill opacity=0.00] (0,0) rectangle (505.89,505.89);
\begin{scope}
\path[clip] (  0.00,  0.00) rectangle (505.89,505.89);
\definecolor{drawColor}{RGB}{255,255,255}

\path[draw=drawColor,line width= 0.4pt,line join=round,line cap=round] ( 65.18, 75.85) circle (  2.25);
\end{scope}
\begin{scope}
\path[clip] (  0.00,  0.00) rectangle (505.89,505.89);
\definecolor{drawColor}{RGB}{0,0,0}

\path[draw=drawColor,line width= 0.4pt,line join=round,line cap=round] (115.12, 61.20) -- (414.77, 61.20);

\path[draw=drawColor,line width= 0.4pt,line join=round,line cap=round] (115.12, 61.20) -- (115.12, 55.20);

\path[draw=drawColor,line width= 0.4pt,line join=round,line cap=round] (215.00, 61.20) -- (215.00, 55.20);

\path[draw=drawColor,line width= 0.4pt,line join=round,line cap=round] (314.89, 61.20) -- (314.89, 55.20);

\path[draw=drawColor,line width= 0.4pt,line join=round,line cap=round] (414.77, 61.20) -- (414.77, 55.20);

\node[text=drawColor,anchor=base,inner sep=0pt, outer sep=0pt, scale=  1.00] at (115.12, 39.60) {0.4};

\node[text=drawColor,anchor=base,inner sep=0pt, outer sep=0pt, scale=  1.00] at (215.00, 39.60) {0.6};

\node[text=drawColor,anchor=base,inner sep=0pt, outer sep=0pt, scale=  1.00] at (314.89, 39.60) {0.8};

\node[text=drawColor,anchor=base,inner sep=0pt, outer sep=0pt, scale=  1.00] at (414.77, 39.60) {1.0};

\path[draw=drawColor,line width= 0.4pt,line join=round,line cap=round] ( 49.20, 75.85) -- ( 49.20,413.41);

\path[draw=drawColor,line width= 0.4pt,line join=round,line cap=round] ( 49.20, 75.85) -- ( 43.20, 75.85);

\path[draw=drawColor,line width= 0.4pt,line join=round,line cap=round] ( 49.20,160.24) -- ( 43.20,160.24);

\path[draw=drawColor,line width= 0.4pt,line join=round,line cap=round] ( 49.20,244.63) -- ( 43.20,244.63);

\path[draw=drawColor,line width= 0.4pt,line join=round,line cap=round] ( 49.20,329.02) -- ( 43.20,329.02);

\path[draw=drawColor,line width= 0.4pt,line join=round,line cap=round] ( 49.20,413.41) -- ( 43.20,413.41);

\node[text=drawColor,rotate= 90.00,anchor=base,inner sep=0pt, outer sep=0pt, scale=  1.00] at ( 34.80, 75.85) {0.00};

\node[text=drawColor,rotate= 90.00,anchor=base,inner sep=0pt, outer sep=0pt, scale=  1.00] at ( 34.80,160.24) {0.05};

\node[text=drawColor,rotate= 90.00,anchor=base,inner sep=0pt, outer sep=0pt, scale=  1.00] at ( 34.80,244.63) {0.10};

\node[text=drawColor,rotate= 90.00,anchor=base,inner sep=0pt, outer sep=0pt, scale=  1.00] at ( 34.80,329.02) {0.15};

\node[text=drawColor,rotate= 90.00,anchor=base,inner sep=0pt, outer sep=0pt, scale=  1.00] at ( 34.80,413.41) {0.20};

\path[draw=drawColor,line width= 0.4pt,line join=round,line cap=round] ( 49.20, 61.20) --
	(480.69, 61.20) --
	(480.69,456.69) --
	( 49.20,456.69) --
	( 49.20, 61.20);
\end{scope}
\begin{scope}
\path[clip] (  0.00,  0.00) rectangle (505.89,505.89);
\definecolor{drawColor}{RGB}{0,0,0}

\node[text=drawColor,anchor=base,inner sep=0pt, outer sep=0pt, scale=  1.00] at (264.94, 15.60) {d};

\node[text=drawColor,rotate= 90.00,anchor=base,inner sep=0pt, outer sep=0pt, scale=  1.00] at ( 10.80,258.94) {MSE};

\path[draw=drawColor,line width= 0.4pt,line join=round,line cap=round] ( 65.18,352.96) circle (  2.25);

\path[draw=drawColor,line width= 0.4pt,line join=round,line cap=round] (115.12,344.62) circle (  2.25);

\path[draw=drawColor,line width= 0.4pt,line join=round,line cap=round] (165.06,334.28) circle (  2.25);

\path[draw=drawColor,line width= 0.4pt,line join=round,line cap=round] (215.00,321.23) circle (  2.25);

\path[draw=drawColor,line width= 0.4pt,line join=round,line cap=round] (264.94,307.55) circle (  2.25);

\path[draw=drawColor,line width= 0.4pt,line join=round,line cap=round] (314.89,293.79) circle (  2.25);

\path[draw=drawColor,line width= 0.4pt,line join=round,line cap=round] (364.83,278.66) circle (  2.25);

\path[draw=drawColor,line width= 0.4pt,line join=round,line cap=round] (414.77,263.29) circle (  2.25);

\path[draw=drawColor,line width= 0.4pt,line join=round,line cap=round] (464.71,247.80) circle (  2.25);

\path[draw=drawColor,line width= 0.4pt,dash pattern=on 4pt off 4pt ,line join=round,line cap=round] ( 65.18,352.96) --
	(115.12,344.62) --
	(165.06,334.28) --
	(215.00,321.23) --
	(264.94,307.55) --
	(314.89,293.79) --
	(364.83,278.66) --
	(414.77,263.29) --
	(464.71,247.80);

\path[draw=drawColor,line width= 0.4pt,line join=round,line cap=round] ( 65.18,429.87) circle (  2.25);

\path[draw=drawColor,line width= 0.4pt,line join=round,line cap=round] (115.12,416.24) circle (  2.25);

\path[draw=drawColor,line width= 0.4pt,line join=round,line cap=round] (165.06,401.46) circle (  2.25);

\path[draw=drawColor,line width= 0.4pt,line join=round,line cap=round] (215.00,383.12) circle (  2.25);

\path[draw=drawColor,line width= 0.4pt,line join=round,line cap=round] (264.94,361.82) circle (  2.25);

\path[draw=drawColor,line width= 0.4pt,line join=round,line cap=round] (314.89,340.74) circle (  2.25);

\path[draw=drawColor,line width= 0.4pt,line join=round,line cap=round] (364.83,318.60) circle (  2.25);

\path[draw=drawColor,line width= 0.4pt,line join=round,line cap=round] (414.77,295.01) circle (  2.25);

\path[draw=drawColor,line width= 0.4pt,line join=round,line cap=round] (464.71,271.13) circle (  2.25);

\path[draw=drawColor,line width= 0.4pt,dash pattern=on 1pt off 3pt ,line join=round,line cap=round] ( 65.18,429.87) --
	(115.12,416.24) --
	(165.06,401.46) --
	(215.00,383.12) --
	(264.94,361.82) --
	(314.89,340.74) --
	(364.83,318.60) --
	(414.77,295.01) --
	(464.71,271.13);

\path[draw=drawColor,line width= 0.4pt,line join=round,line cap=round] ( 65.18,315.27) --
	( 68.21,310.02) --
	( 62.15,310.02) --
	( 65.18,315.27);

\path[draw=drawColor,line width= 0.4pt,line join=round,line cap=round] (115.12,283.83) --
	(118.15,278.58) --
	(112.09,278.58) --
	(115.12,283.83);

\path[draw=drawColor,line width= 0.4pt,line join=round,line cap=round] (165.06,250.27) --
	(168.09,245.03) --
	(162.03,245.03) --
	(165.06,250.27);

\path[draw=drawColor,line width= 0.4pt,line join=round,line cap=round] (215.00,218.10) --
	(218.03,212.85) --
	(211.97,212.85) --
	(215.00,218.10);

\path[draw=drawColor,line width= 0.4pt,line join=round,line cap=round] (264.94,190.22) --
	(267.98,184.97) --
	(261.91,184.97) --
	(264.94,190.22);

\path[draw=drawColor,line width= 0.4pt,line join=round,line cap=round] (314.89,167.01) --
	(317.92,161.76) --
	(311.86,161.76) --
	(314.89,167.01);

\path[draw=drawColor,line width= 0.4pt,line join=round,line cap=round] (364.83,148.75) --
	(367.86,143.50) --
	(361.80,143.50) --
	(364.83,148.75);

\path[draw=drawColor,line width= 0.4pt,line join=round,line cap=round] (414.77,134.12) --
	(417.80,128.87) --
	(411.74,128.87) --
	(414.77,134.12);

\path[draw=drawColor,line width= 0.4pt,line join=round,line cap=round] (464.71,123.37) --
	(467.74,118.13) --
	(461.68,118.13) --
	(464.71,123.37);

\path[draw=drawColor,line width= 0.4pt,dash pattern=on 4pt off 4pt ,line join=round,line cap=round] ( 65.18,311.77) --
	(115.12,280.33) --
	(165.06,246.78) --
	(215.00,214.60) --
	(264.94,186.72) --
	(314.89,163.51) --
	(364.83,145.25) --
	(414.77,130.62) --
	(464.71,119.88);

\path[draw=drawColor,line width= 0.4pt,line join=round,line cap=round] ( 65.18,424.06) --
	( 68.21,418.81) --
	( 62.15,418.81) --
	( 65.18,424.06);

\path[draw=drawColor,line width= 0.4pt,line join=round,line cap=round] (115.12,367.50) --
	(118.15,362.25) --
	(112.09,362.25) --
	(115.12,367.50);

\path[draw=drawColor,line width= 0.4pt,line join=round,line cap=round] (165.06,308.42) --
	(168.09,303.17) --
	(162.03,303.17) --
	(165.06,308.42);

\path[draw=drawColor,line width= 0.4pt,line join=round,line cap=round] (215.00,252.72) --
	(218.03,247.48) --
	(211.97,247.48) --
	(215.00,252.72);

\path[draw=drawColor,line width= 0.4pt,line join=round,line cap=round] (264.94,205.72) --
	(267.98,200.47) --
	(261.91,200.47) --
	(264.94,205.72);

\path[draw=drawColor,line width= 0.4pt,line join=round,line cap=round] (314.89,166.61) --
	(317.92,161.36) --
	(311.86,161.36) --
	(314.89,166.61);

\path[draw=drawColor,line width= 0.4pt,line join=round,line cap=round] (364.83,137.19) --
	(367.86,131.94) --
	(361.80,131.94) --
	(364.83,137.19);

\path[draw=drawColor,line width= 0.4pt,line join=round,line cap=round] (414.77,116.22) --
	(417.80,110.97) --
	(411.74,110.97) --
	(414.77,116.22);

\path[draw=drawColor,line width= 0.4pt,line join=round,line cap=round] (464.71,102.43) --
	(467.74, 97.18) --
	(461.68, 97.18) --
	(464.71,102.43);

\path[draw=drawColor,line width= 0.4pt,dash pattern=on 1pt off 3pt ,line join=round,line cap=round] ( 65.18,420.56) --
	(115.12,364.00) --
	(165.06,304.92) --
	(215.00,249.23) --
	(264.94,202.22) --
	(314.89,163.11) --
	(364.83,133.69) --
	(414.77,112.72) --
	(464.71, 98.93);

\path[draw=drawColor,line width= 0.4pt,line join=round,line cap=round] (244.25,484.24) circle (  2.25);

\path[draw=drawColor,line width= 0.4pt,line join=round,line cap=round] (271.69,487.74) --
	(274.73,482.49) --
	(268.66,482.49) --
	(271.69,487.74);

\node[text=drawColor,anchor=base west,inner sep=0pt, outer sep=0pt, scale=  1.00] at (253.25,480.80) {i)};

\node[text=drawColor,anchor=base west,inner sep=0pt, outer sep=0pt, scale=  1.00] at (280.69,480.80) {ii)};

\path[draw=drawColor,line width= 0.4pt,dash pattern=on 4pt off 4pt ,line join=round,line cap=round] (223.75,464.46) -- (241.75,464.46);

\path[draw=drawColor,line width= 0.4pt,dash pattern=on 1pt off 3pt ,line join=round,line cap=round] (271.70,464.46) -- (289.69,464.46);

\node[text=drawColor,anchor=base west,inner sep=0pt, outer sep=0pt, scale=  1.00] at (250.75,461.02) {i)};

\node[text=drawColor,anchor=base west,inner sep=0pt, outer sep=0pt, scale=  1.00] at (298.69,461.02) {vi)};
\end{scope}
\end{tikzpicture}}
	}
	\subfloat[$\tau=0.5$]{
		\resizebox{0.48\textwidth}{!}{\input{stdNormal_MSE_GaK_lambda=50pc.tex}}
	}
	
	\caption{MSE for Cusum and weighted Cusum change-point estimator based on $M=10^5$ simulations of $N$   i.i. $\mathcal{N}(0,1)$-distributed random variables with change of size $d=0.4$ at $\tau$}
	\label{fig:Density_CPE_Finite}
\end{figure}

\begin{figure}[H]
	\subfloat[$N=50$, $\tau=0.15$]{
	\includegraphics[scale=0.35]{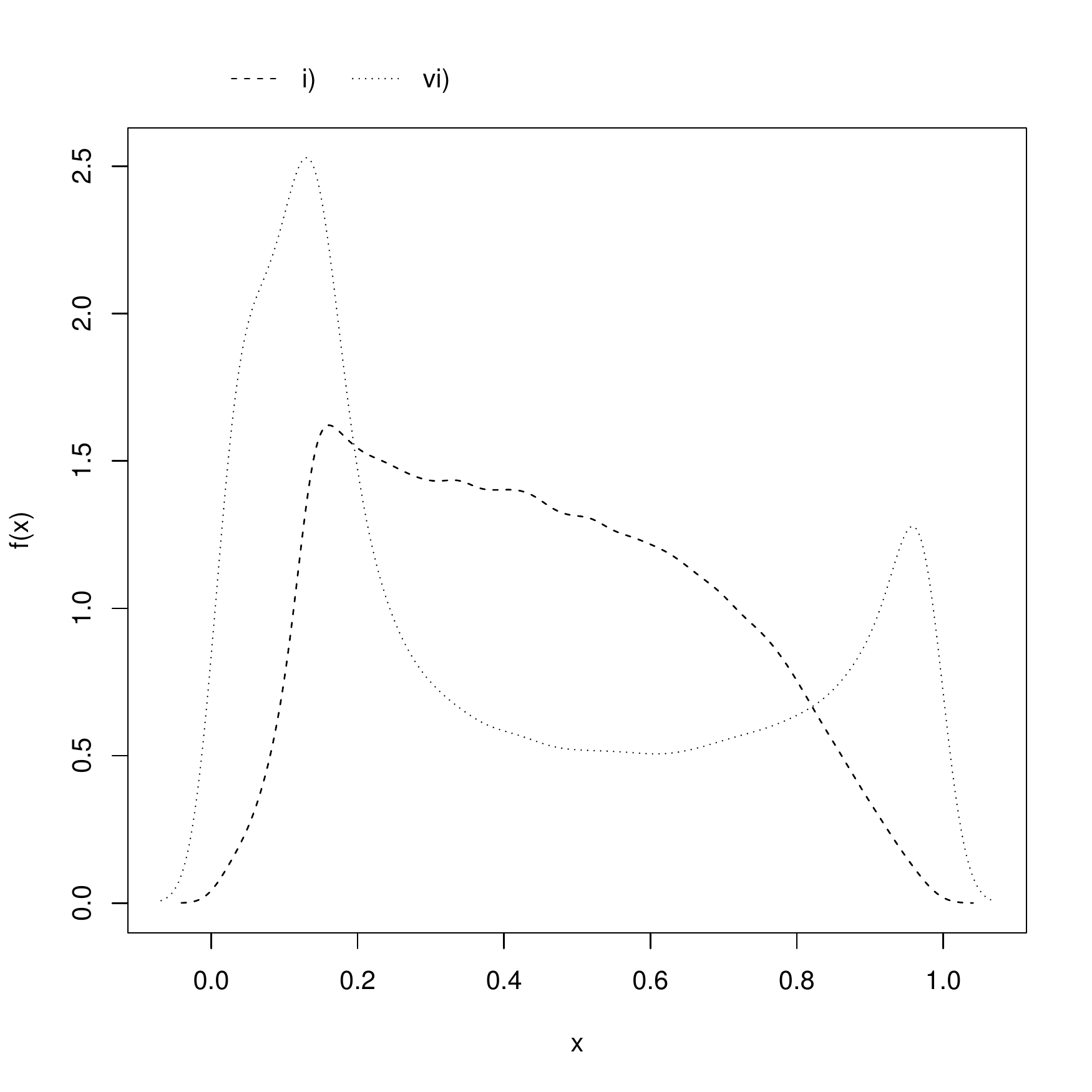}
}
\subfloat[$N=50$, $\tau=0.5$]{
	\includegraphics[scale=0.35]{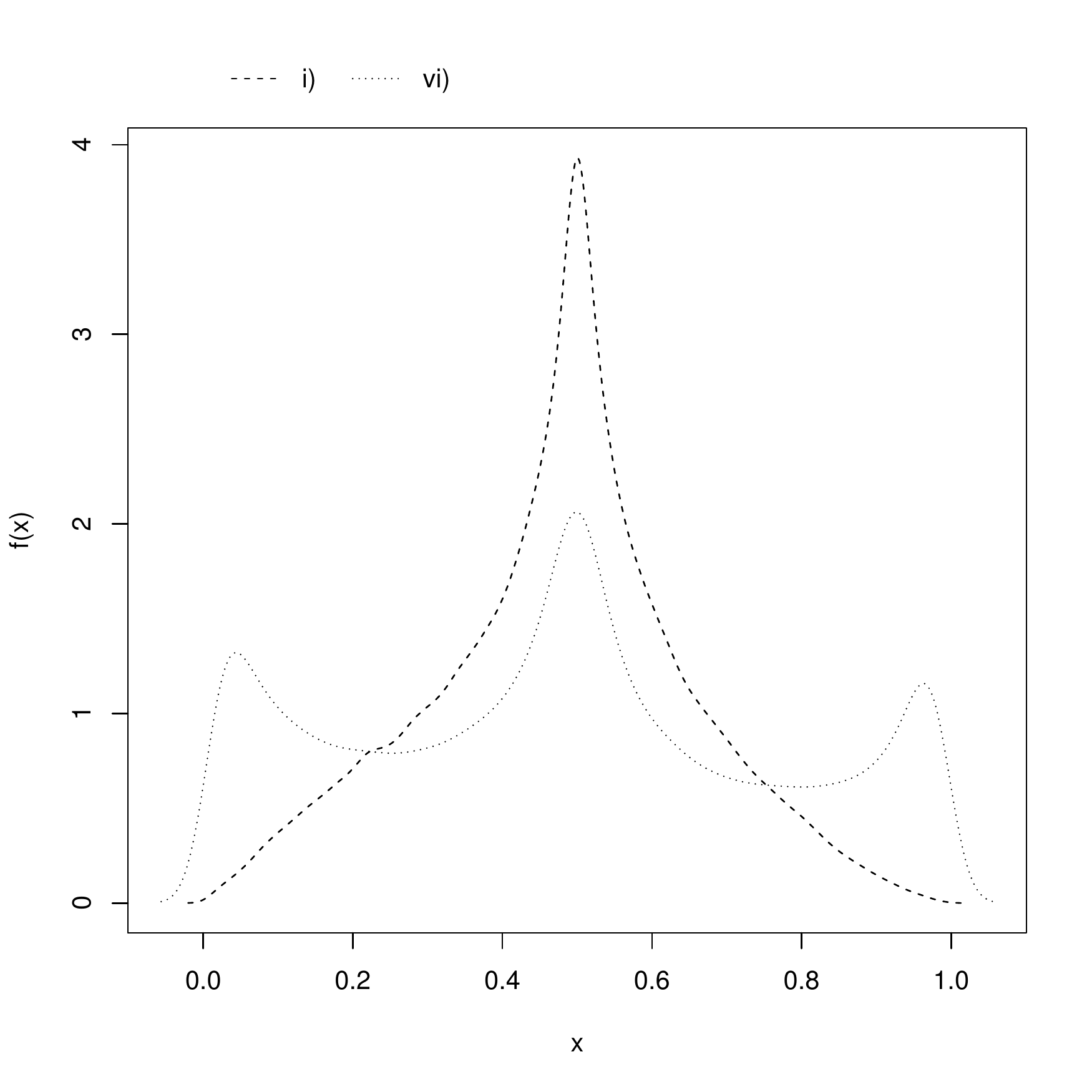}
}\\
	\subfloat[$N=250$, $\tau=0.15$]{
	\includegraphics[scale=0.35]{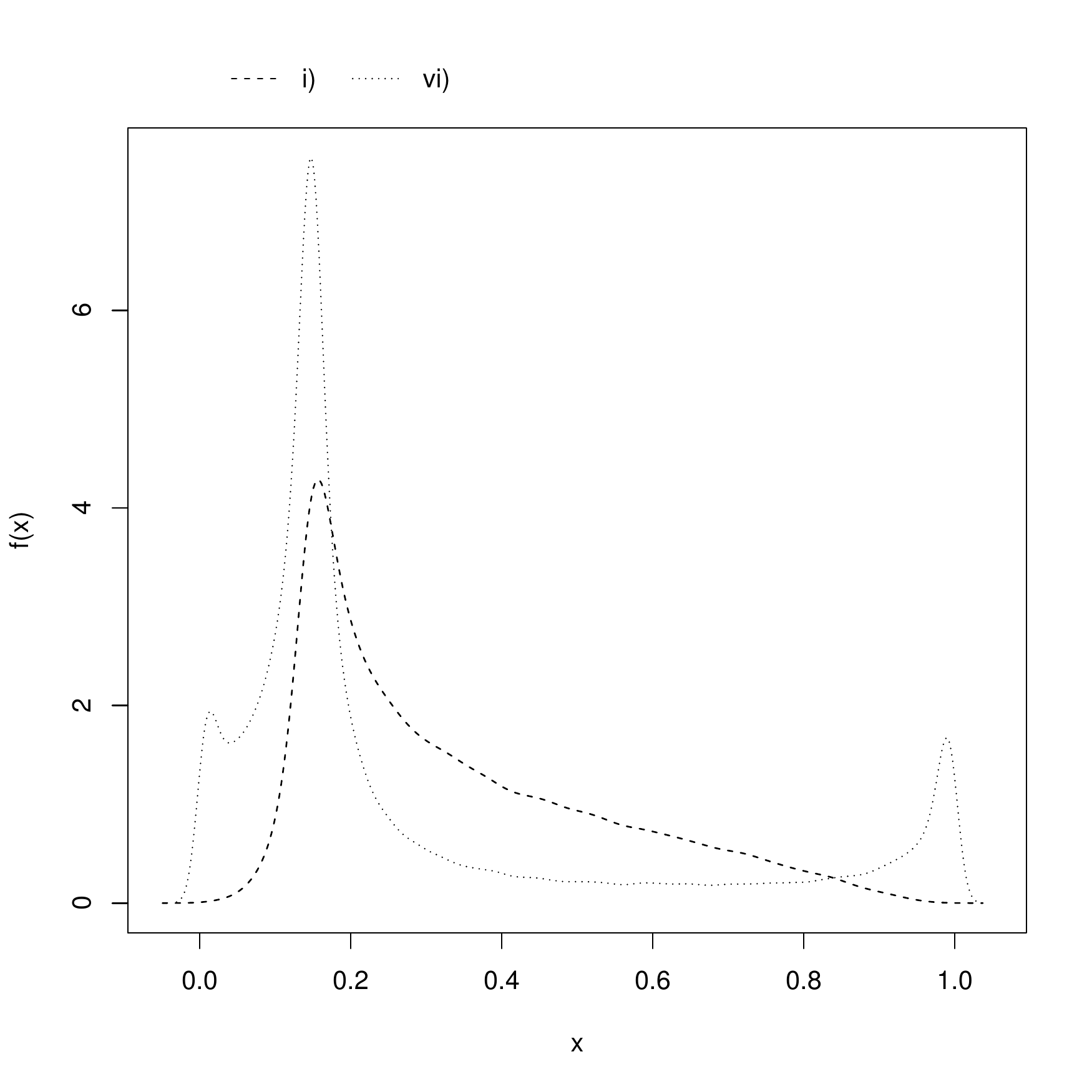}
}
\subfloat[$N=250$, $\tau=0.5$]{
	\includegraphics[scale=0.35]{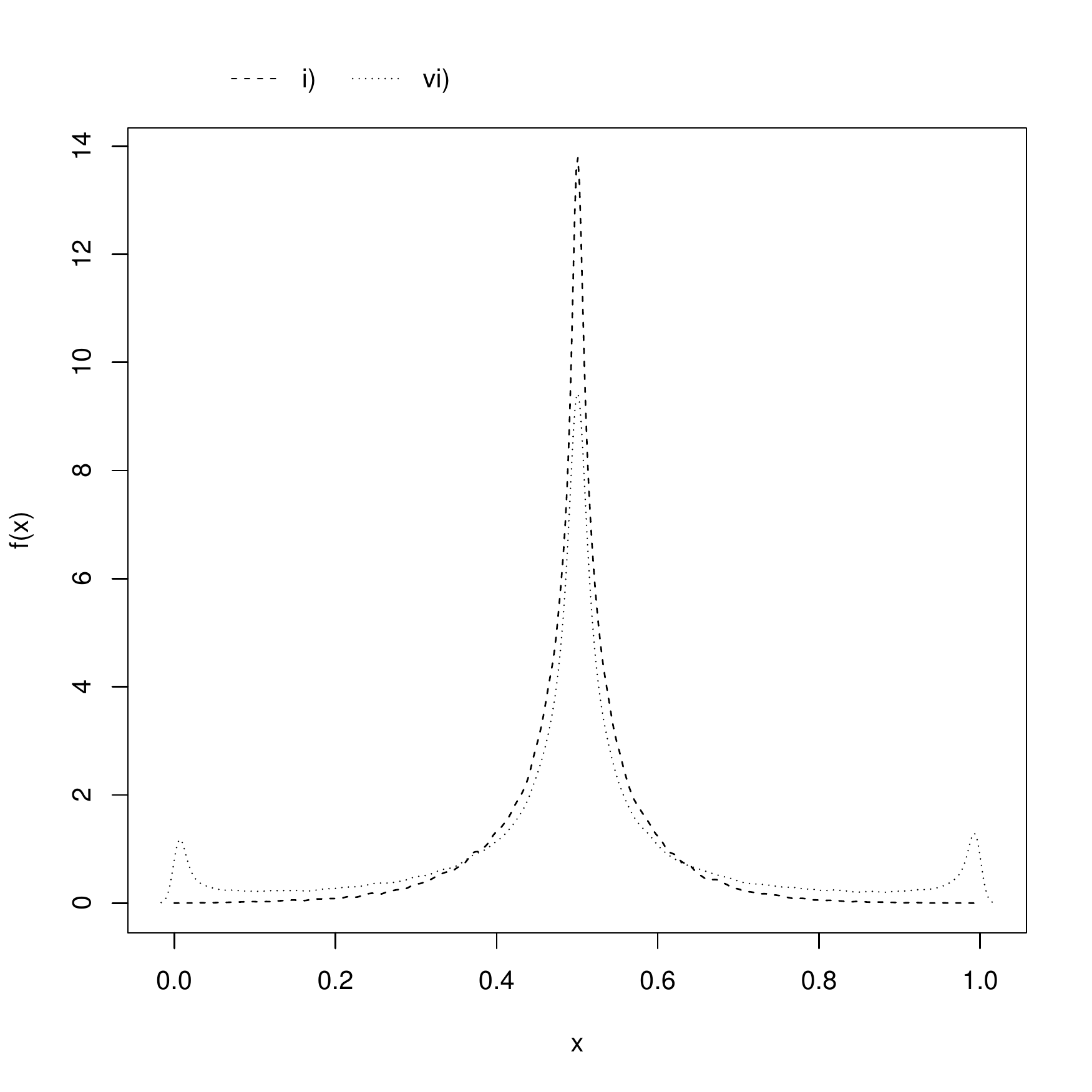}
}

\caption{rescaled density of Cusum and weighted Cusum change-point estimator based on $M=10^5$ simulations of $N$ i.i. $Exp(1)$-distributed random variables with change of size $d=0.4$ at $\tau$}
	\label{fig:Density_CPE_Finite}
\end{figure}

\begin{figure}[H]
	\subfloat[$\tau=0.15$]{
		\resizebox{0.48\textwidth}{!}{
\begin{tikzpicture}[x=1pt,y=1pt]
\definecolor{fillColor}{RGB}{255,255,255}
\path[use as bounding box,fill=fillColor,fill opacity=0.00] (0,0) rectangle (505.89,505.89);
\begin{scope}
\path[clip] (  0.00,  0.00) rectangle (505.89,505.89);
\definecolor{drawColor}{RGB}{255,255,255}

\path[draw=drawColor,line width= 0.4pt,line join=round,line cap=round] ( 65.18, 75.85) circle (  2.25);
\end{scope}
\begin{scope}
\path[clip] (  0.00,  0.00) rectangle (505.89,505.89);
\definecolor{drawColor}{RGB}{0,0,0}

\path[draw=drawColor,line width= 0.4pt,line join=round,line cap=round] (115.12, 61.20) -- (414.77, 61.20);

\path[draw=drawColor,line width= 0.4pt,line join=round,line cap=round] (115.12, 61.20) -- (115.12, 55.20);

\path[draw=drawColor,line width= 0.4pt,line join=round,line cap=round] (215.00, 61.20) -- (215.00, 55.20);

\path[draw=drawColor,line width= 0.4pt,line join=round,line cap=round] (314.89, 61.20) -- (314.89, 55.20);

\path[draw=drawColor,line width= 0.4pt,line join=round,line cap=round] (414.77, 61.20) -- (414.77, 55.20);

\node[text=drawColor,anchor=base,inner sep=0pt, outer sep=0pt, scale=  1.00] at (115.12, 39.60) {0.4};

\node[text=drawColor,anchor=base,inner sep=0pt, outer sep=0pt, scale=  1.00] at (215.00, 39.60) {0.6};

\node[text=drawColor,anchor=base,inner sep=0pt, outer sep=0pt, scale=  1.00] at (314.89, 39.60) {0.8};

\node[text=drawColor,anchor=base,inner sep=0pt, outer sep=0pt, scale=  1.00] at (414.77, 39.60) {1.0};

\path[draw=drawColor,line width= 0.4pt,line join=round,line cap=round] ( 49.20, 75.85) -- ( 49.20,364.09);

\path[draw=drawColor,line width= 0.4pt,line join=round,line cap=round] ( 49.20, 75.85) -- ( 43.20, 75.85);

\path[draw=drawColor,line width= 0.4pt,line join=round,line cap=round] ( 49.20,171.93) -- ( 43.20,171.93);

\path[draw=drawColor,line width= 0.4pt,line join=round,line cap=round] ( 49.20,268.01) -- ( 43.20,268.01);

\path[draw=drawColor,line width= 0.4pt,line join=round,line cap=round] ( 49.20,364.09) -- ( 43.20,364.09);

\node[text=drawColor,rotate= 90.00,anchor=base,inner sep=0pt, outer sep=0pt, scale=  1.00] at ( 34.80, 75.85) {0.00};

\node[text=drawColor,rotate= 90.00,anchor=base,inner sep=0pt, outer sep=0pt, scale=  1.00] at ( 34.80,171.93) {0.05};

\node[text=drawColor,rotate= 90.00,anchor=base,inner sep=0pt, outer sep=0pt, scale=  1.00] at ( 34.80,268.01) {0.10};

\node[text=drawColor,rotate= 90.00,anchor=base,inner sep=0pt, outer sep=0pt, scale=  1.00] at ( 34.80,364.09) {0.15};

\path[draw=drawColor,line width= 0.4pt,line join=round,line cap=round] ( 49.20, 61.20) --
	(480.69, 61.20) --
	(480.69,456.69) --
	( 49.20,456.69) --
	( 49.20, 61.20);
\end{scope}
\begin{scope}
\path[clip] (  0.00,  0.00) rectangle (505.89,505.89);
\definecolor{drawColor}{RGB}{0,0,0}

\node[text=drawColor,anchor=base,inner sep=0pt, outer sep=0pt, scale=  1.00] at (264.94, 15.60) {d};

\node[text=drawColor,rotate= 90.00,anchor=base,inner sep=0pt, outer sep=0pt, scale=  1.00] at ( 10.80,258.94) {MSE};

\path[draw=drawColor,line width= 0.4pt,line join=round,line cap=round] ( 65.18,378.36) circle (  2.25);

\path[draw=drawColor,line width= 0.4pt,line join=round,line cap=round] (115.12,363.54) circle (  2.25);

\path[draw=drawColor,line width= 0.4pt,line join=round,line cap=round] (165.06,346.81) circle (  2.25);

\path[draw=drawColor,line width= 0.4pt,line join=round,line cap=round] (215.00,329.30) circle (  2.25);

\path[draw=drawColor,line width= 0.4pt,line join=round,line cap=round] (264.94,311.84) circle (  2.25);

\path[draw=drawColor,line width= 0.4pt,line join=round,line cap=round] (314.89,294.20) circle (  2.25);

\path[draw=drawColor,line width= 0.4pt,line join=round,line cap=round] (364.83,276.66) circle (  2.25);

\path[draw=drawColor,line width= 0.4pt,line join=round,line cap=round] (414.77,260.86) circle (  2.25);

\path[draw=drawColor,line width= 0.4pt,line join=round,line cap=round] (464.71,245.48) circle (  2.25);

\path[draw=drawColor,line width= 0.4pt,dash pattern=on 4pt off 4pt ,line join=round,line cap=round] ( 65.18,378.36) --
	(115.12,363.54) --
	(165.06,346.81) --
	(215.00,329.30) --
	(264.94,311.84) --
	(314.89,294.20) --
	(364.83,276.66) --
	(414.77,260.86) --
	(464.71,245.48);

\path[draw=drawColor,line width= 0.4pt,line join=round,line cap=round] ( 65.18,432.25) circle (  2.25);

\path[draw=drawColor,line width= 0.4pt,line join=round,line cap=round] (115.12,408.48) circle (  2.25);

\path[draw=drawColor,line width= 0.4pt,line join=round,line cap=round] (165.06,383.90) circle (  2.25);

\path[draw=drawColor,line width= 0.4pt,line join=round,line cap=round] (215.00,359.23) circle (  2.25);

\path[draw=drawColor,line width= 0.4pt,line join=round,line cap=round] (264.94,335.61) circle (  2.25);

\path[draw=drawColor,line width= 0.4pt,line join=round,line cap=round] (314.89,313.07) circle (  2.25);

\path[draw=drawColor,line width= 0.4pt,line join=round,line cap=round] (364.83,290.15) circle (  2.25);

\path[draw=drawColor,line width= 0.4pt,line join=round,line cap=round] (414.77,269.95) circle (  2.25);

\path[draw=drawColor,line width= 0.4pt,line join=round,line cap=round] (464.71,250.26) circle (  2.25);

\path[draw=drawColor,line width= 0.4pt,dash pattern=on 1pt off 3pt ,line join=round,line cap=round] ( 65.18,432.25) --
	(115.12,408.48) --
	(165.06,383.90) --
	(215.00,359.23) --
	(264.94,335.61) --
	(314.89,313.07) --
	(364.83,290.15) --
	(414.77,269.95) --
	(464.71,250.26);

\path[draw=drawColor,line width= 0.4pt,line join=round,line cap=round] ( 65.18,334.79) --
	( 68.21,329.54) --
	( 62.15,329.54) --
	( 65.18,334.79);

\path[draw=drawColor,line width= 0.4pt,line join=round,line cap=round] (115.12,297.79) --
	(118.15,292.54) --
	(112.09,292.54) --
	(115.12,297.79);

\path[draw=drawColor,line width= 0.4pt,line join=round,line cap=round] (165.06,260.37) --
	(168.09,255.12) --
	(162.03,255.12) --
	(165.06,260.37);

\path[draw=drawColor,line width= 0.4pt,line join=round,line cap=round] (215.00,227.36) --
	(218.03,222.11) --
	(211.97,222.11) --
	(215.00,227.36);

\path[draw=drawColor,line width= 0.4pt,line join=round,line cap=round] (264.94,198.30) --
	(267.98,193.05) --
	(261.91,193.05) --
	(264.94,198.30);

\path[draw=drawColor,line width= 0.4pt,line join=round,line cap=round] (314.89,174.68) --
	(317.92,169.44) --
	(311.86,169.44) --
	(314.89,174.68);

\path[draw=drawColor,line width= 0.4pt,line join=round,line cap=round] (364.83,156.07) --
	(367.86,150.82) --
	(361.80,150.82) --
	(364.83,156.07);

\path[draw=drawColor,line width= 0.4pt,line join=round,line cap=round] (414.77,141.40) --
	(417.80,136.15) --
	(411.74,136.15) --
	(414.77,141.40);

\path[draw=drawColor,line width= 0.4pt,line join=round,line cap=round] (464.71,129.86) --
	(467.74,124.61) --
	(461.68,124.61) --
	(464.71,129.86);

\path[draw=drawColor,line width= 0.4pt,dash pattern=on 4pt off 4pt ,line join=round,line cap=round] ( 65.18,331.29) --
	(115.12,294.29) --
	(165.06,256.87) --
	(215.00,223.86) --
	(264.94,194.80) --
	(314.89,171.19) --
	(364.83,152.57) --
	(414.77,137.90) --
	(464.71,126.36);

\path[draw=drawColor,line width= 0.4pt,line join=round,line cap=round] ( 65.18,422.77) --
	( 68.21,417.52) --
	( 62.15,417.52) --
	( 65.18,422.77);

\path[draw=drawColor,line width= 0.4pt,line join=round,line cap=round] (115.12,362.75) --
	(118.15,357.50) --
	(112.09,357.50) --
	(115.12,362.75);

\path[draw=drawColor,line width= 0.4pt,line join=round,line cap=round] (165.06,306.42) --
	(168.09,301.17) --
	(162.03,301.17) --
	(165.06,306.42);

\path[draw=drawColor,line width= 0.4pt,line join=round,line cap=round] (215.00,257.04) --
	(218.03,251.80) --
	(211.97,251.80) --
	(215.00,257.04);

\path[draw=drawColor,line width= 0.4pt,line join=round,line cap=round] (264.94,214.86) --
	(267.98,209.61) --
	(261.91,209.61) --
	(264.94,214.86);

\path[draw=drawColor,line width= 0.4pt,line join=round,line cap=round] (314.89,181.85) --
	(317.92,176.60) --
	(311.86,176.60) --
	(314.89,181.85);

\path[draw=drawColor,line width= 0.4pt,line join=round,line cap=round] (364.83,155.38) --
	(367.86,150.13) --
	(361.80,150.13) --
	(364.83,155.38);

\path[draw=drawColor,line width= 0.4pt,line join=round,line cap=round] (414.77,134.04) --
	(417.80,128.79) --
	(411.74,128.79) --
	(414.77,134.04);

\path[draw=drawColor,line width= 0.4pt,line join=round,line cap=round] (464.71,118.17) --
	(467.74,112.93) --
	(461.68,112.93) --
	(464.71,118.17);

\path[draw=drawColor,line width= 0.4pt,dash pattern=on 1pt off 3pt ,line join=round,line cap=round] ( 65.18,419.27) --
	(115.12,359.25) --
	(165.06,302.92) --
	(215.00,253.55) --
	(264.94,211.36) --
	(314.89,178.35) --
	(364.83,151.88) --
	(414.77,130.54) --
	(464.71,114.68);

\path[draw=drawColor,line width= 0.4pt,line join=round,line cap=round] (244.25,484.24) circle (  2.25);

\path[draw=drawColor,line width= 0.4pt,line join=round,line cap=round] (271.69,487.74) --
	(274.73,482.49) --
	(268.66,482.49) --
	(271.69,487.74);

\node[text=drawColor,anchor=base west,inner sep=0pt, outer sep=0pt, scale=  1.00] at (253.25,480.80) {i)};

\node[text=drawColor,anchor=base west,inner sep=0pt, outer sep=0pt, scale=  1.00] at (280.69,480.80) {ii)};

\path[draw=drawColor,line width= 0.4pt,dash pattern=on 4pt off 4pt ,line join=round,line cap=round] (223.75,464.46) -- (241.75,464.46);

\path[draw=drawColor,line width= 0.4pt,dash pattern=on 1pt off 3pt ,line join=round,line cap=round] (271.70,464.46) -- (289.69,464.46);

\node[text=drawColor,anchor=base west,inner sep=0pt, outer sep=0pt, scale=  1.00] at (250.75,461.02) {i)};

\node[text=drawColor,anchor=base west,inner sep=0pt, outer sep=0pt, scale=  1.00] at (298.69,461.02) {vi)};
\end{scope}
\end{tikzpicture}}
	}
	\subfloat[$\tau=0.5$]{
		\resizebox{0.48\textwidth}{!}{
\begin{tikzpicture}[x=1pt,y=1pt]
\definecolor{fillColor}{RGB}{255,255,255}
\path[use as bounding box,fill=fillColor,fill opacity=0.00] (0,0) rectangle (505.89,505.89);
\begin{scope}
\path[clip] (  0.00,  0.00) rectangle (505.89,505.89);
\definecolor{drawColor}{RGB}{255,255,255}

\path[draw=drawColor,line width= 0.4pt,line join=round,line cap=round] ( 65.18, 75.85) circle (  2.25);
\end{scope}
\begin{scope}
\path[clip] (  0.00,  0.00) rectangle (505.89,505.89);
\definecolor{drawColor}{RGB}{0,0,0}

\path[draw=drawColor,line width= 0.4pt,line join=round,line cap=round] (115.12, 61.20) -- (414.77, 61.20);

\path[draw=drawColor,line width= 0.4pt,line join=round,line cap=round] (115.12, 61.20) -- (115.12, 55.20);

\path[draw=drawColor,line width= 0.4pt,line join=round,line cap=round] (215.00, 61.20) -- (215.00, 55.20);

\path[draw=drawColor,line width= 0.4pt,line join=round,line cap=round] (314.89, 61.20) -- (314.89, 55.20);

\path[draw=drawColor,line width= 0.4pt,line join=round,line cap=round] (414.77, 61.20) -- (414.77, 55.20);

\node[text=drawColor,anchor=base,inner sep=0pt, outer sep=0pt, scale=  1.00] at (115.12, 39.60) {0.4};

\node[text=drawColor,anchor=base,inner sep=0pt, outer sep=0pt, scale=  1.00] at (215.00, 39.60) {0.6};

\node[text=drawColor,anchor=base,inner sep=0pt, outer sep=0pt, scale=  1.00] at (314.89, 39.60) {0.8};

\node[text=drawColor,anchor=base,inner sep=0pt, outer sep=0pt, scale=  1.00] at (414.77, 39.60) {1.0};

\path[draw=drawColor,line width= 0.4pt,line join=round,line cap=round] ( 49.20, 75.85) -- ( 49.20,395.93);

\path[draw=drawColor,line width= 0.4pt,line join=round,line cap=round] ( 49.20, 75.85) -- ( 43.20, 75.85);

\path[draw=drawColor,line width= 0.4pt,line join=round,line cap=round] ( 49.20,155.87) -- ( 43.20,155.87);

\path[draw=drawColor,line width= 0.4pt,line join=round,line cap=round] ( 49.20,235.89) -- ( 43.20,235.89);

\path[draw=drawColor,line width= 0.4pt,line join=round,line cap=round] ( 49.20,315.91) -- ( 43.20,315.91);

\path[draw=drawColor,line width= 0.4pt,line join=round,line cap=round] ( 49.20,395.93) -- ( 43.20,395.93);

\node[text=drawColor,rotate= 90.00,anchor=base,inner sep=0pt, outer sep=0pt, scale=  1.00] at ( 34.80, 75.85) {0.00};

\node[text=drawColor,rotate= 90.00,anchor=base,inner sep=0pt, outer sep=0pt, scale=  1.00] at ( 34.80,155.87) {0.02};

\node[text=drawColor,rotate= 90.00,anchor=base,inner sep=0pt, outer sep=0pt, scale=  1.00] at ( 34.80,235.89) {0.04};

\node[text=drawColor,rotate= 90.00,anchor=base,inner sep=0pt, outer sep=0pt, scale=  1.00] at ( 34.80,315.91) {0.06};

\node[text=drawColor,rotate= 90.00,anchor=base,inner sep=0pt, outer sep=0pt, scale=  1.00] at ( 34.80,395.93) {0.08};

\path[draw=drawColor,line width= 0.4pt,line join=round,line cap=round] ( 49.20, 61.20) --
	(480.69, 61.20) --
	(480.69,456.69) --
	( 49.20,456.69) --
	( 49.20, 61.20);
\end{scope}
\begin{scope}
\path[clip] (  0.00,  0.00) rectangle (505.89,505.89);
\definecolor{drawColor}{RGB}{0,0,0}

\node[text=drawColor,anchor=base,inner sep=0pt, outer sep=0pt, scale=  1.00] at (264.94, 15.60) {d};

\node[text=drawColor,rotate= 90.00,anchor=base,inner sep=0pt, outer sep=0pt, scale=  1.00] at ( 10.80,258.94) {MSE};

\path[draw=drawColor,line width= 0.4pt,line join=round,line cap=round] ( 65.18,250.80) circle (  2.25);

\path[draw=drawColor,line width= 0.4pt,line join=round,line cap=round] (115.12,236.38) circle (  2.25);

\path[draw=drawColor,line width= 0.4pt,line join=round,line cap=round] (165.06,220.03) circle (  2.25);

\path[draw=drawColor,line width= 0.4pt,line join=round,line cap=round] (215.00,202.84) circle (  2.25);

\path[draw=drawColor,line width= 0.4pt,line join=round,line cap=round] (264.94,185.39) circle (  2.25);

\path[draw=drawColor,line width= 0.4pt,line join=round,line cap=round] (314.89,168.92) circle (  2.25);

\path[draw=drawColor,line width= 0.4pt,line join=round,line cap=round] (364.83,154.19) circle (  2.25);

\path[draw=drawColor,line width= 0.4pt,line join=round,line cap=round] (414.77,141.05) circle (  2.25);

\path[draw=drawColor,line width= 0.4pt,line join=round,line cap=round] (464.71,129.58) circle (  2.25);

\path[draw=drawColor,line width= 0.4pt,dash pattern=on 4pt off 4pt ,line join=round,line cap=round] ( 65.18,250.80) --
	(115.12,236.38) --
	(165.06,220.03) --
	(215.00,202.84) --
	(264.94,185.39) --
	(314.89,168.92) --
	(364.83,154.19) --
	(414.77,141.05) --
	(464.71,129.58);

\path[draw=drawColor,line width= 0.4pt,line join=round,line cap=round] ( 65.18,406.98) circle (  2.25);

\path[draw=drawColor,line width= 0.4pt,line join=round,line cap=round] (115.12,388.47) circle (  2.25);

\path[draw=drawColor,line width= 0.4pt,line join=round,line cap=round] (165.06,365.69) circle (  2.25);

\path[draw=drawColor,line width= 0.4pt,line join=round,line cap=round] (215.00,341.80) circle (  2.25);

\path[draw=drawColor,line width= 0.4pt,line join=round,line cap=round] (264.94,317.00) circle (  2.25);

\path[draw=drawColor,line width= 0.4pt,line join=round,line cap=round] (314.89,291.18) circle (  2.25);

\path[draw=drawColor,line width= 0.4pt,line join=round,line cap=round] (364.83,266.72) circle (  2.25);

\path[draw=drawColor,line width= 0.4pt,line join=round,line cap=round] (414.77,244.12) circle (  2.25);

\path[draw=drawColor,line width= 0.4pt,line join=round,line cap=round] (464.71,221.88) circle (  2.25);

\path[draw=drawColor,line width= 0.4pt,dash pattern=on 1pt off 3pt ,line join=round,line cap=round] ( 65.18,406.98) --
	(115.12,388.47) --
	(165.06,365.69) --
	(215.00,341.80) --
	(264.94,317.00) --
	(314.89,291.18) --
	(364.83,266.72) --
	(414.77,244.12) --
	(464.71,221.88);

\path[draw=drawColor,line width= 0.4pt,line join=round,line cap=round] ( 65.18,193.44) --
	( 68.21,188.19) --
	( 62.15,188.19) --
	( 65.18,193.44);

\path[draw=drawColor,line width= 0.4pt,line join=round,line cap=round] (115.12,157.83) --
	(118.15,152.58) --
	(112.09,152.58) --
	(115.12,157.83);

\path[draw=drawColor,line width= 0.4pt,line join=round,line cap=round] (165.06,131.06) --
	(168.09,125.81) --
	(162.03,125.81) --
	(165.06,131.06);

\path[draw=drawColor,line width= 0.4pt,line join=round,line cap=round] (215.00,112.75) --
	(218.03,107.51) --
	(211.97,107.51) --
	(215.00,112.75);

\path[draw=drawColor,line width= 0.4pt,line join=round,line cap=round] (264.94,100.93) --
	(267.98, 95.68) --
	(261.91, 95.68) --
	(264.94,100.93);

\path[draw=drawColor,line width= 0.4pt,line join=round,line cap=round] (314.89, 93.61) --
	(317.92, 88.36) --
	(311.86, 88.36) --
	(314.89, 93.61);

\path[draw=drawColor,line width= 0.4pt,line join=round,line cap=round] (364.83, 89.09) --
	(367.86, 83.84) --
	(361.80, 83.84) --
	(364.83, 89.09);

\path[draw=drawColor,line width= 0.4pt,line join=round,line cap=round] (414.77, 86.20) --
	(417.80, 80.95) --
	(411.74, 80.95) --
	(414.77, 86.20);

\path[draw=drawColor,line width= 0.4pt,line join=round,line cap=round] (464.71, 84.29) --
	(467.74, 79.04) --
	(461.68, 79.04) --
	(464.71, 84.29);

\path[draw=drawColor,line width= 0.4pt,dash pattern=on 4pt off 4pt ,line join=round,line cap=round] ( 65.18,189.94) --
	(115.12,154.33) --
	(165.06,127.56) --
	(215.00,109.25) --
	(264.94, 97.43) --
	(314.89, 90.11) --
	(364.83, 85.59) --
	(414.77, 82.70) --
	(464.71, 80.79);

\path[draw=drawColor,line width= 0.4pt,line join=round,line cap=round] ( 65.18,430.96) --
	( 68.21,425.71) --
	( 62.15,425.71) --
	( 65.18,430.96);

\path[draw=drawColor,line width= 0.4pt,line join=round,line cap=round] (115.12,353.60) --
	(118.15,348.36) --
	(112.09,348.36) --
	(115.12,353.60);

\path[draw=drawColor,line width= 0.4pt,line join=round,line cap=round] (165.06,282.32) --
	(168.09,277.08) --
	(162.03,277.08) --
	(165.06,282.32);

\path[draw=drawColor,line width= 0.4pt,line join=round,line cap=round] (215.00,222.71) --
	(218.03,217.46) --
	(211.97,217.46) --
	(215.00,222.71);

\path[draw=drawColor,line width= 0.4pt,line join=round,line cap=round] (264.94,176.17) --
	(267.98,170.92) --
	(261.91,170.92) --
	(264.94,176.17);

\path[draw=drawColor,line width= 0.4pt,line join=round,line cap=round] (314.89,143.60) --
	(317.92,138.35) --
	(311.86,138.35) --
	(314.89,143.60);

\path[draw=drawColor,line width= 0.4pt,line join=round,line cap=round] (364.83,120.89) --
	(367.86,115.64) --
	(361.80,115.64) --
	(364.83,120.89);

\path[draw=drawColor,line width= 0.4pt,line join=round,line cap=round] (414.77,106.17) --
	(417.80,100.92) --
	(411.74,100.92) --
	(414.77,106.17);

\path[draw=drawColor,line width= 0.4pt,line join=round,line cap=round] (464.71, 96.55) --
	(467.74, 91.30) --
	(461.68, 91.30) --
	(464.71, 96.55);

\path[draw=drawColor,line width= 0.4pt,dash pattern=on 1pt off 3pt ,line join=round,line cap=round] ( 65.18,427.46) --
	(115.12,350.10) --
	(165.06,278.83) --
	(215.00,219.21) --
	(264.94,172.67) --
	(314.89,140.10) --
	(364.83,117.39) --
	(414.77,102.67) --
	(464.71, 93.05);

\path[draw=drawColor,line width= 0.4pt,line join=round,line cap=round] (244.25,484.24) circle (  2.25);

\path[draw=drawColor,line width= 0.4pt,line join=round,line cap=round] (271.69,487.74) --
	(274.73,482.49) --
	(268.66,482.49) --
	(271.69,487.74);

\node[text=drawColor,anchor=base west,inner sep=0pt, outer sep=0pt, scale=  1.00] at (253.25,480.80) {50};

\node[text=drawColor,anchor=base west,inner sep=0pt, outer sep=0pt, scale=  1.00] at (280.69,480.80) {250};

\path[draw=drawColor,line width= 0.4pt,dash pattern=on 4pt off 4pt ,line join=round,line cap=round] (223.75,464.46) -- (241.75,464.46);

\path[draw=drawColor,line width= 0.4pt,dash pattern=on 1pt off 3pt ,line join=round,line cap=round] (271.70,464.46) -- (289.69,464.46);

\node[text=drawColor,anchor=base west,inner sep=0pt, outer sep=0pt, scale=  1.00] at (250.75,461.02) {i)};

\node[text=drawColor,anchor=base west,inner sep=0pt, outer sep=0pt, scale=  1.00] at (298.69,461.02) {vi)};
\end{scope}
\end{tikzpicture}}
	}
	
	\caption{MSE for Cusum and weighted Cusum change-point estimator based on $M=10^5$ simulations of $N$   i.i. $Exp(1)$-distributed and $\mathcal{N}(0,1)$-distributed random variables with change of size $d=0.4$ at $\tau$}
	\label{fig:Density_CPE_Finite}
\end{figure}
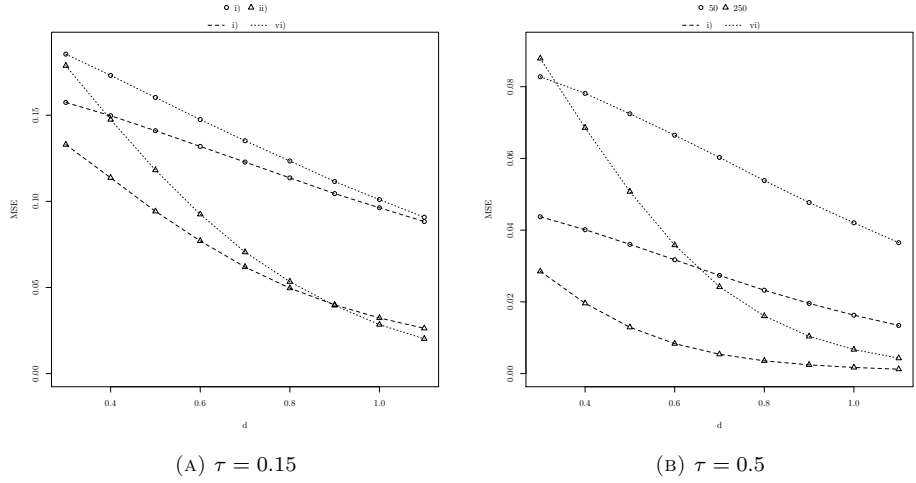
\newpage
\section{data-driven weighted Cusum}\label{Appendix:Esti_data-driven}
~\\
\begin{figure}[H]
	\subfloat[$N=50$, $\tau=0.15$]{
		\includegraphics[scale=0.35]{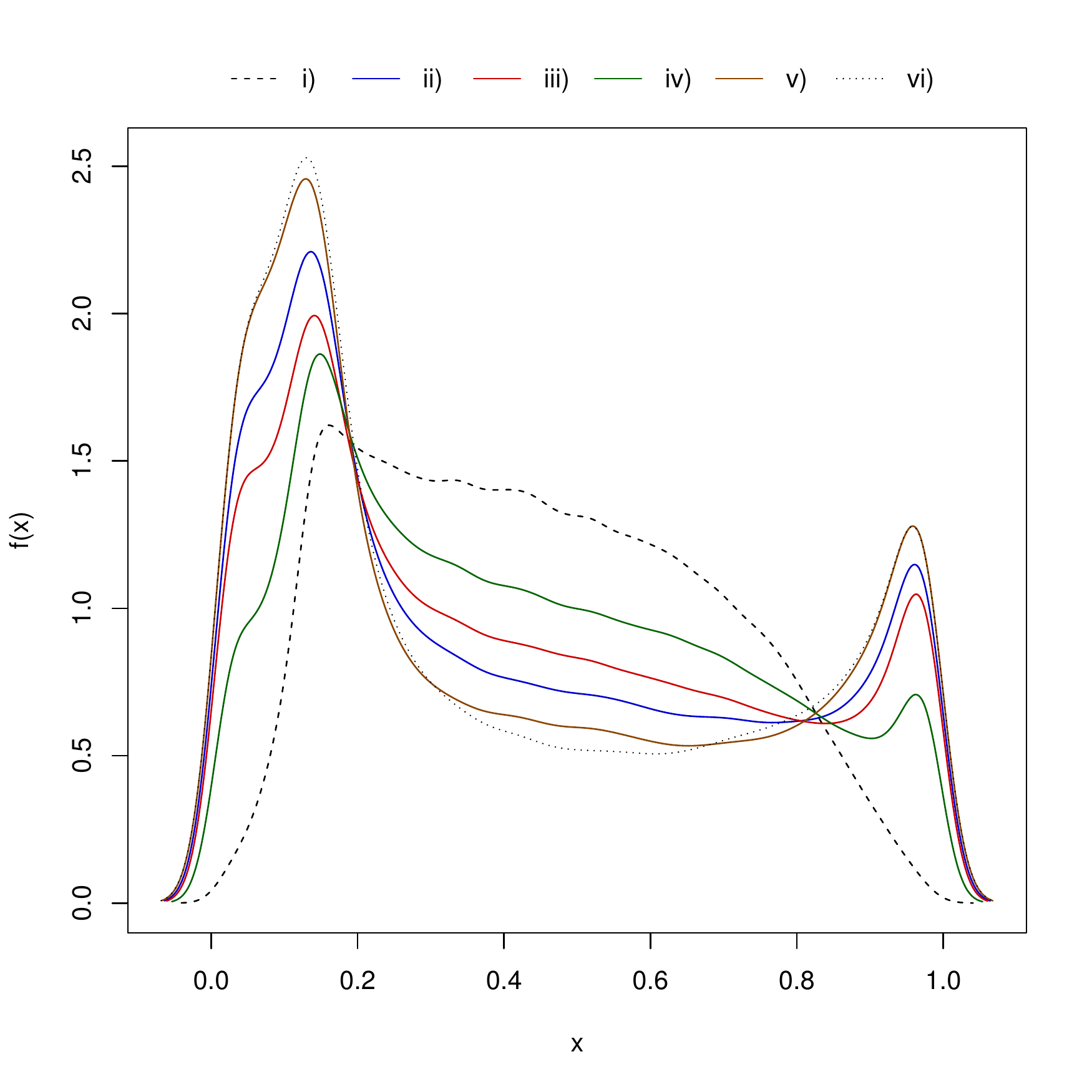}
	}
	\subfloat[$N=50$, $\tau=0.5$]{
		\includegraphics[scale=0.35]{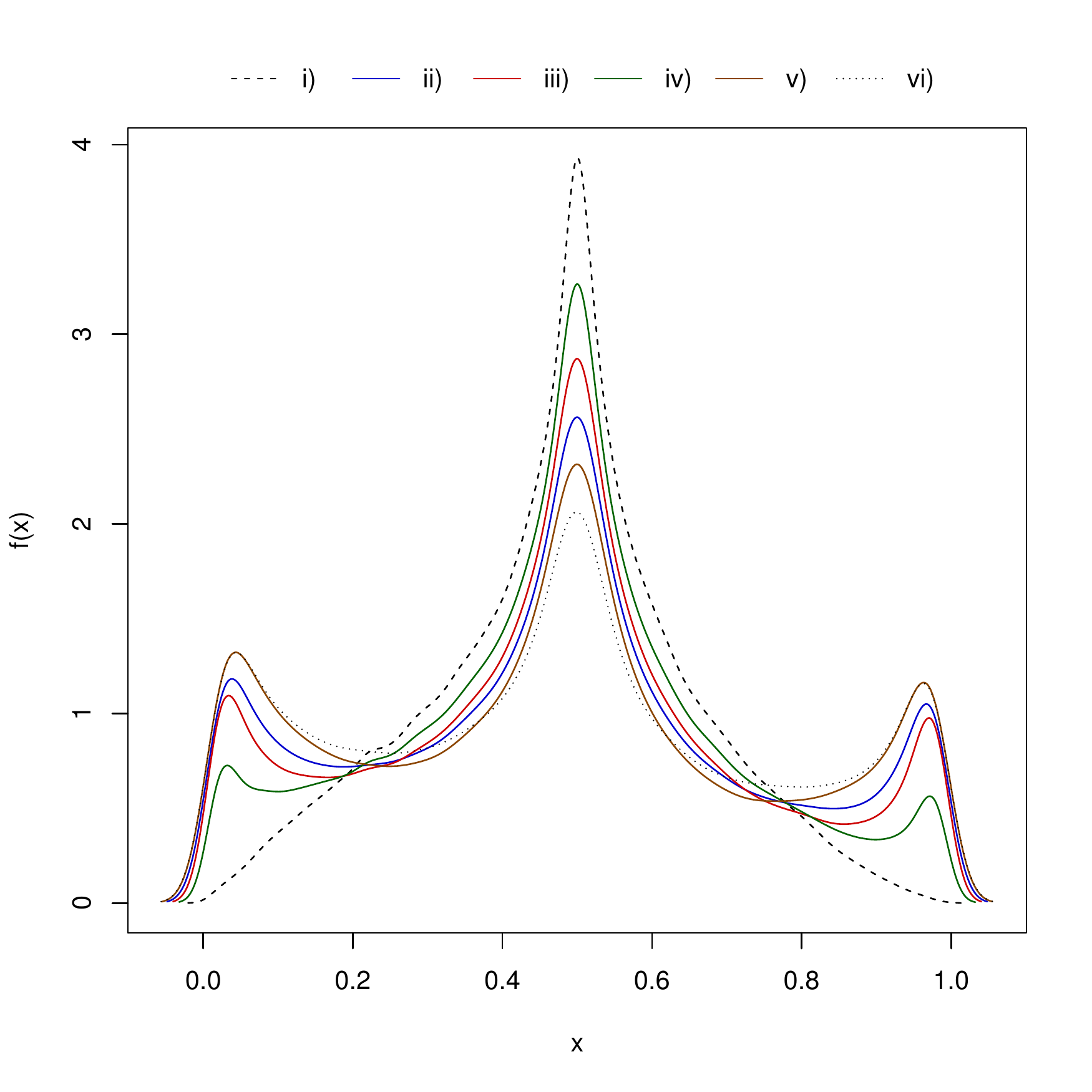}
	}\\
	\subfloat[$N=250$, $\tau=0.15$]{
		\includegraphics[scale=0.35]{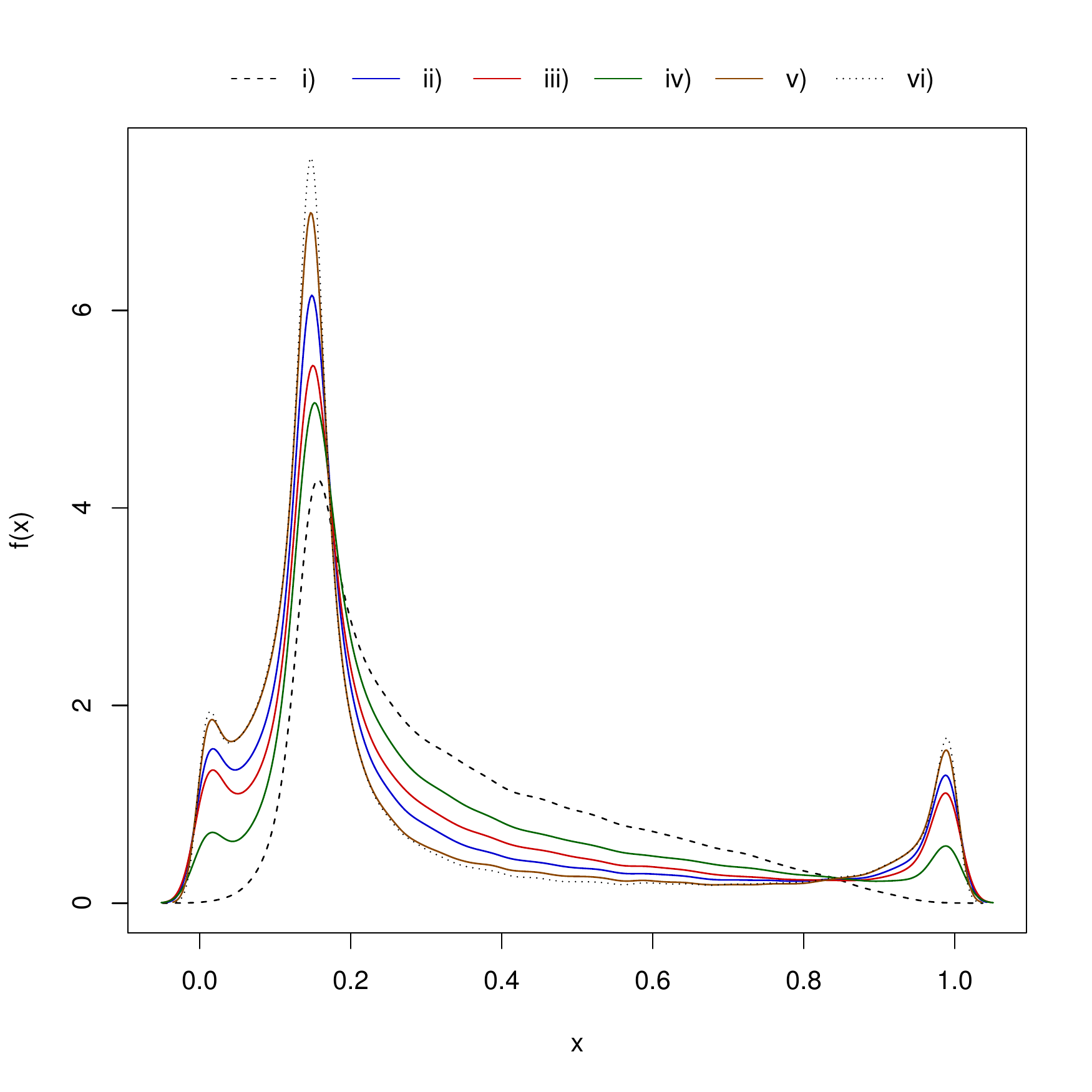}
	}
	\subfloat[$N=250$, $\tau=0.5$]{
		\includegraphics[scale=0.35]{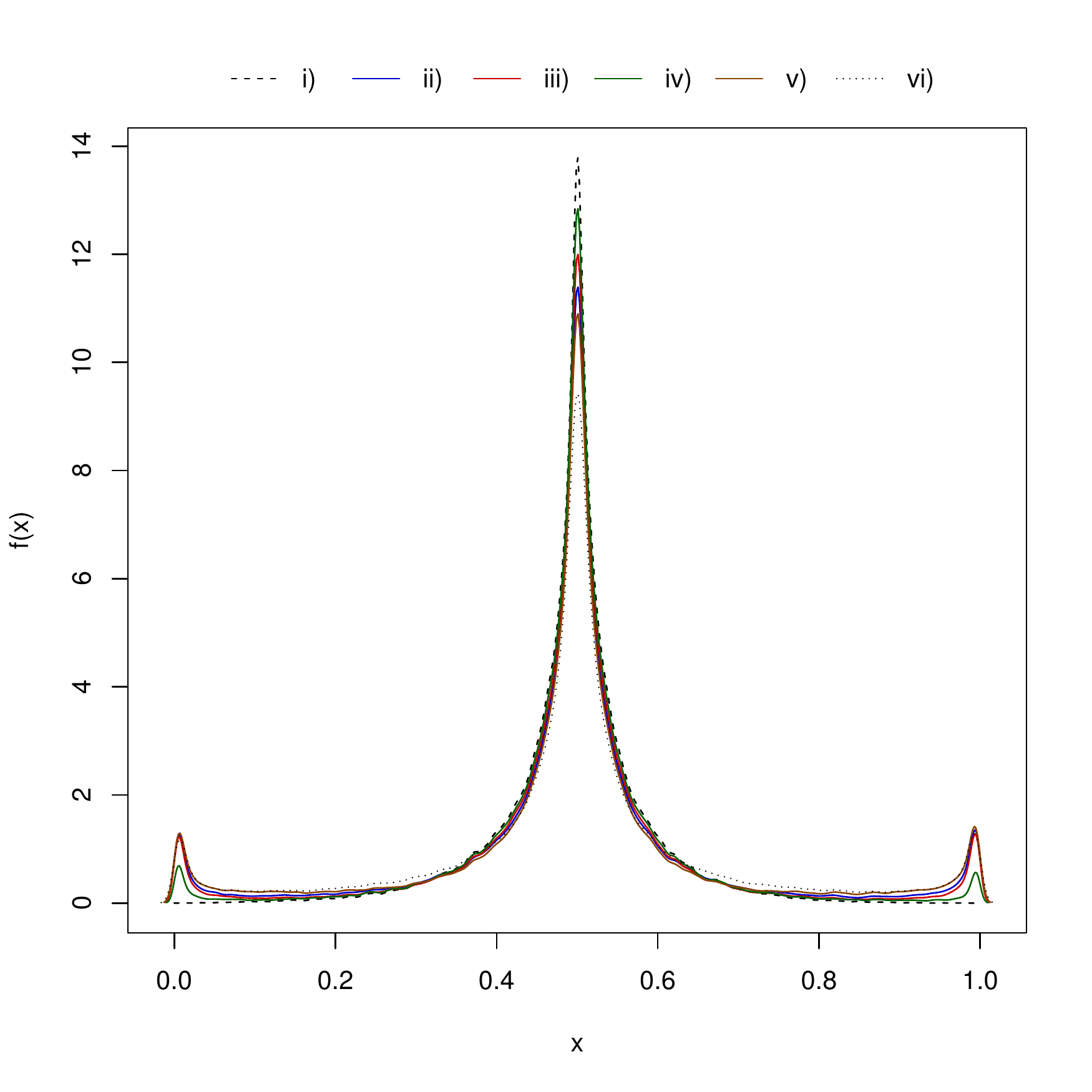}
	}
	\caption{empirical MSE ($M=10^5$ repetitions) of the plug-in change-point estimator for i.i. $Exp(1)$-distributed observations with change of size $d=0.4$ at $\tau=0.15$ (A) and (C) or $\tau=0.5$ (B) and (D) }
\end{figure}
~\\
\begin{figure}[H]
	\subfloat[$\tau=0.15$]{
		\resizebox{0.48\textwidth}{!}{\input{Exp1_MSE_lambda=15pc.tex}}
	}
	\subfloat[$\tau=0.5$]{
		\resizebox{0.48\textwidth}{!}{\input{Exp1_MSE_lambda=50pc.tex}}
	}
	\caption{empirical MSE ($M=10^5$ repetitions) of the plug-in change-point estimator for i.i. $Exp(1)$-distributed observations with change of size $d$ at $\tau=0.15$ (A) or $\tau=0.5$ (B) }
\end{figure}
\begin{figure}[H]
	\subfloat[$N=50$, $\tau=0.15$]{
		\includegraphics[scale=0.35]{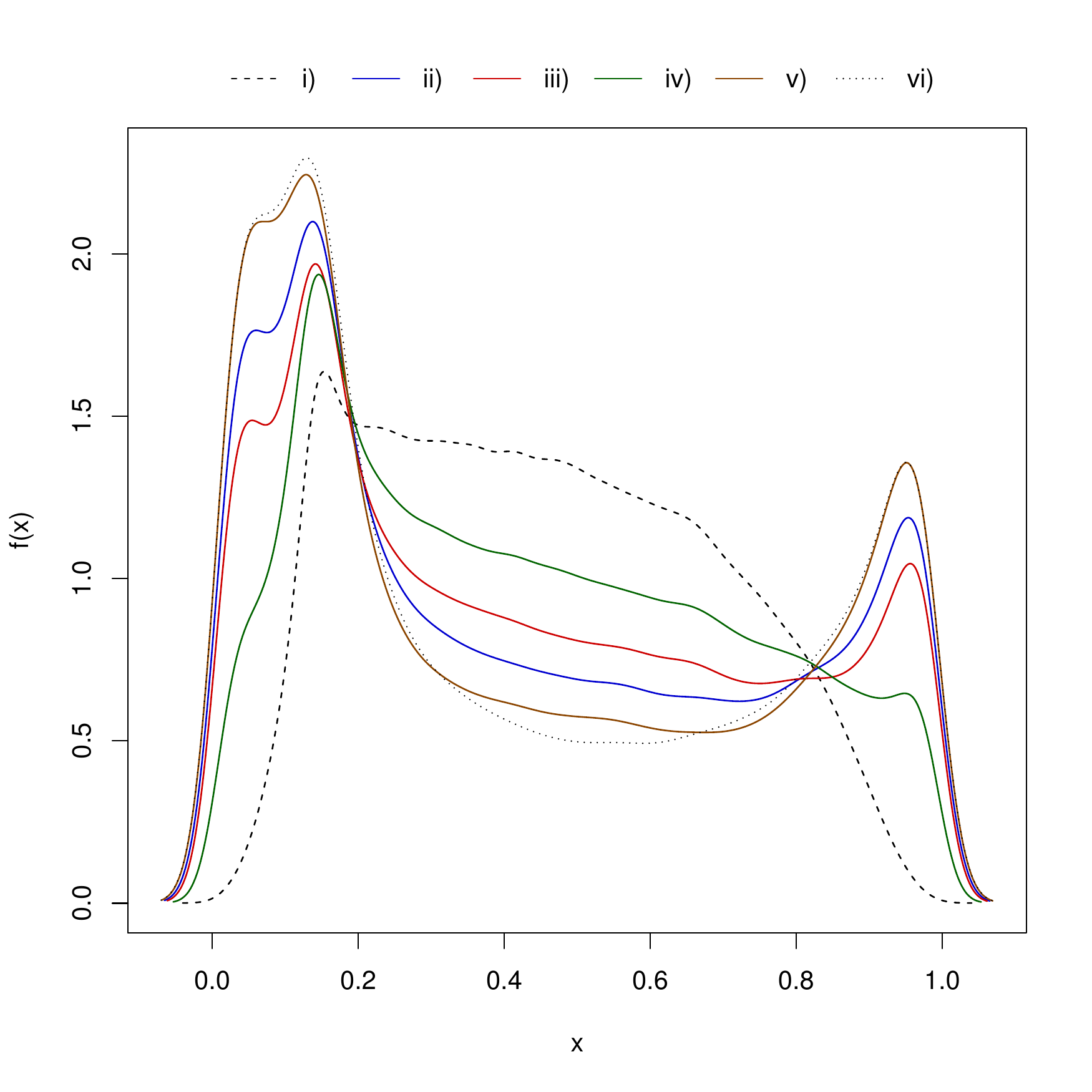}
	}
	\subfloat[$N=50$, $\tau=0.5$]{
		\includegraphics[scale=0.35]{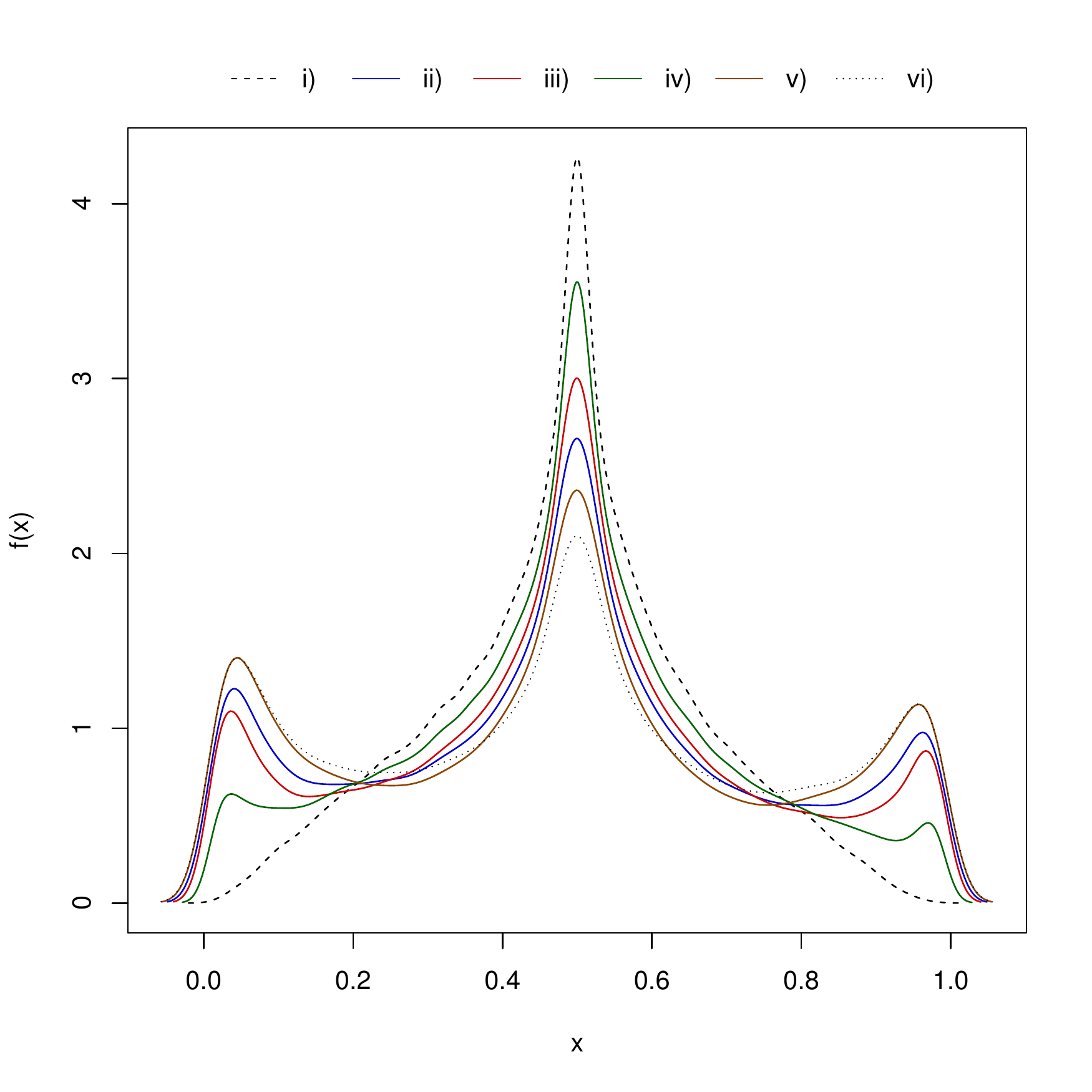}
	}\\
	\subfloat[$N=250$, $\tau=0.15$]{
		\includegraphics[scale=0.35]{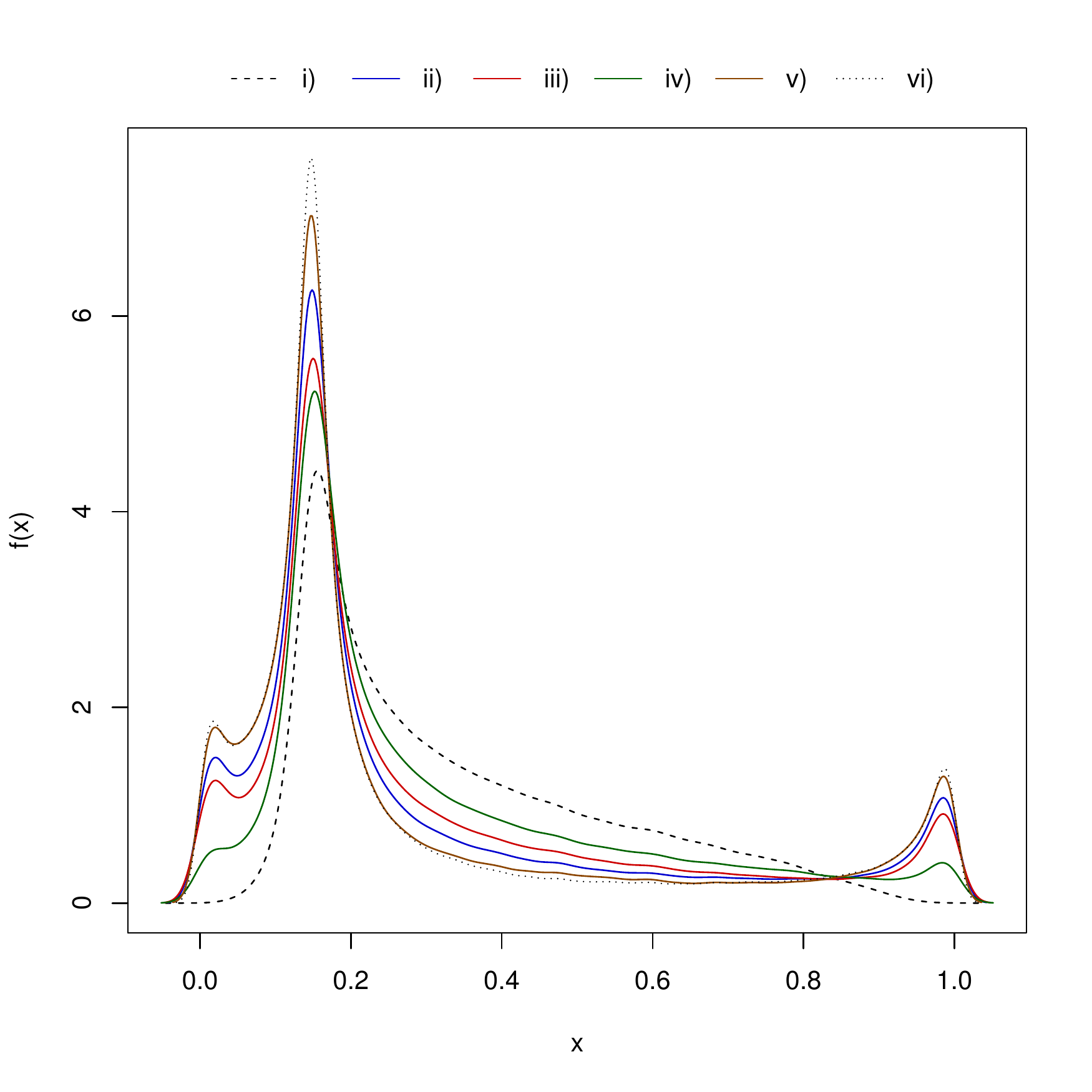}
	}
	\subfloat[$N=250$, $\tau=0.5$]{
		\includegraphics[scale=0.35]{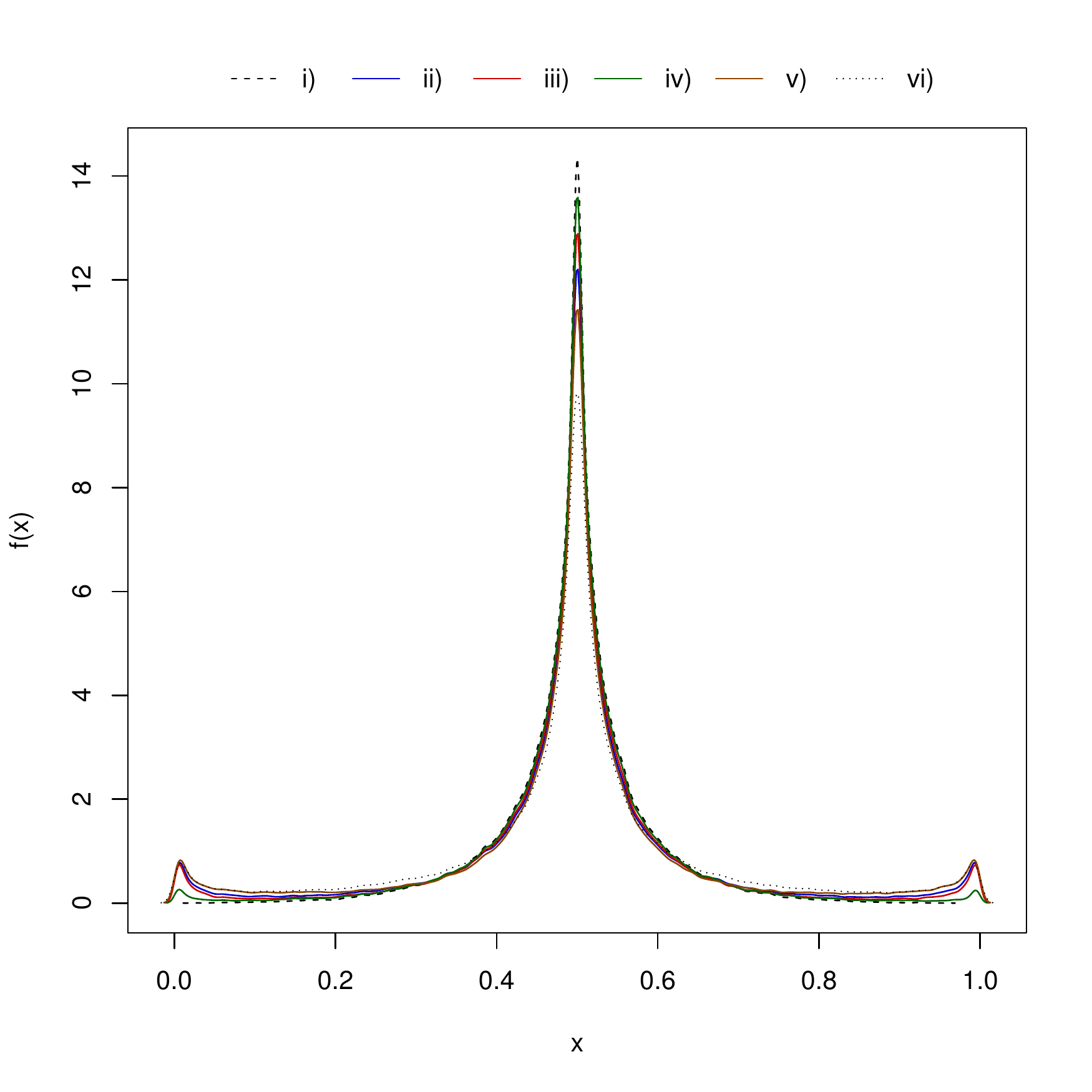}
	}
	\caption{empirical MSE ($M=10^5$ repetitions) of the plug-in change-point estimator for i.i. $Poi(1)$-distributed observations with change of size $d=0.4$ at $\tau=0.15$ (A) and (C) or $\tau=0.5$ (B) and (D) }
\end{figure}
\begin{figure}[H]
	\subfloat[$\tau=0.15$]{
		\resizebox{0.48\textwidth}{!}{\input{Poi1_MSE_lambda=15pc.tex}}
	}
	\subfloat[$\tau=0.5$]{
		\resizebox{0.48\textwidth}{!}{\input{Poi1_MSE_lambda=50pc.tex}}
	}
	\caption{empirical MSE ($M=10^5$ repetitions) of the plug-in change-point estimator for i.i. $Poi([0,1])$-distributed observations with change of size $d$ at $\tau=0.15$ (A) or $\tau=0.5$ (B) }
\end{figure}
\begin{figure}[H]
	\subfloat[$N=50$, $\tau=0.15$]{
		\includegraphics[scale=0.35]{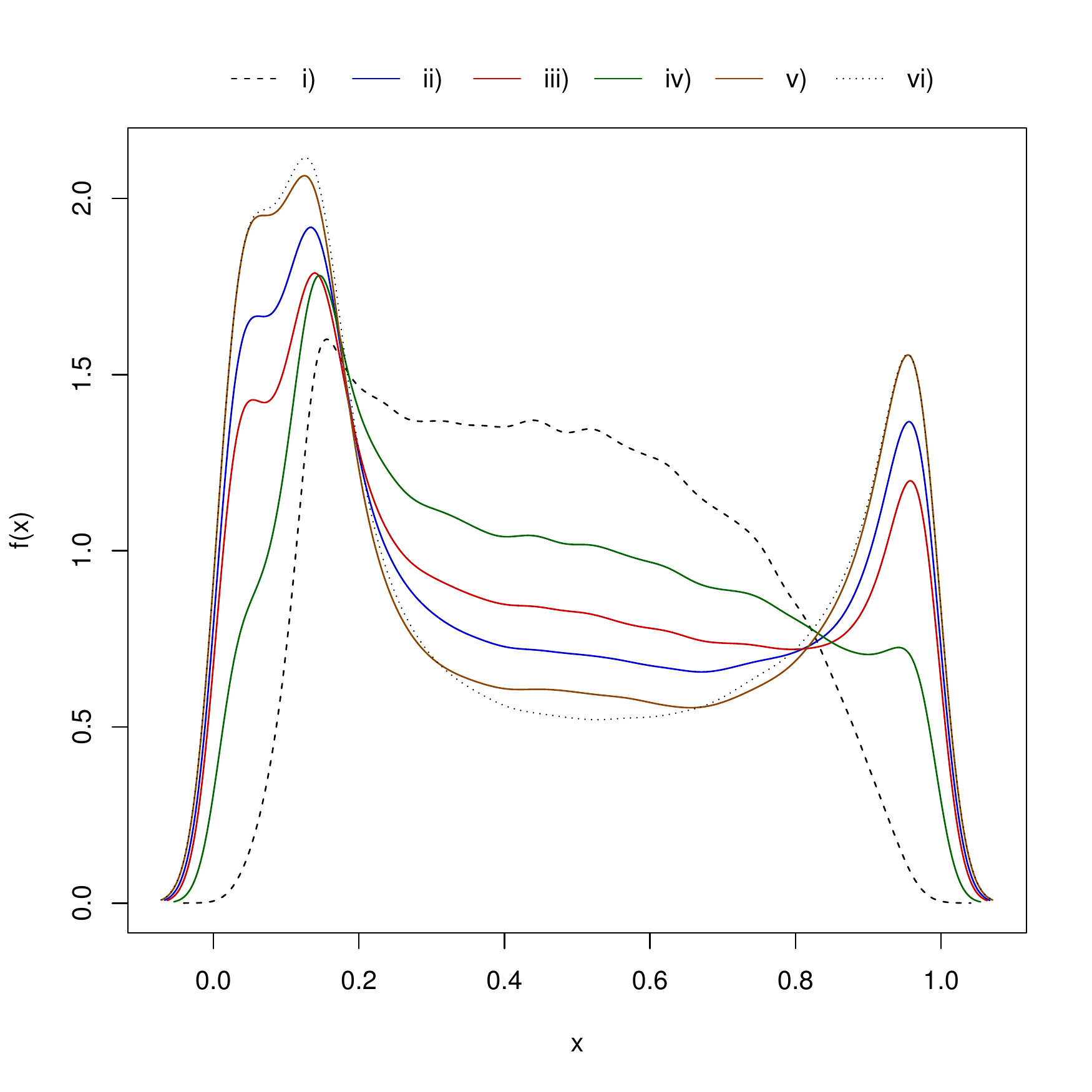}
	}
	\subfloat[$N=50$, $\tau=0.5$]{
		\includegraphics[scale=0.35]{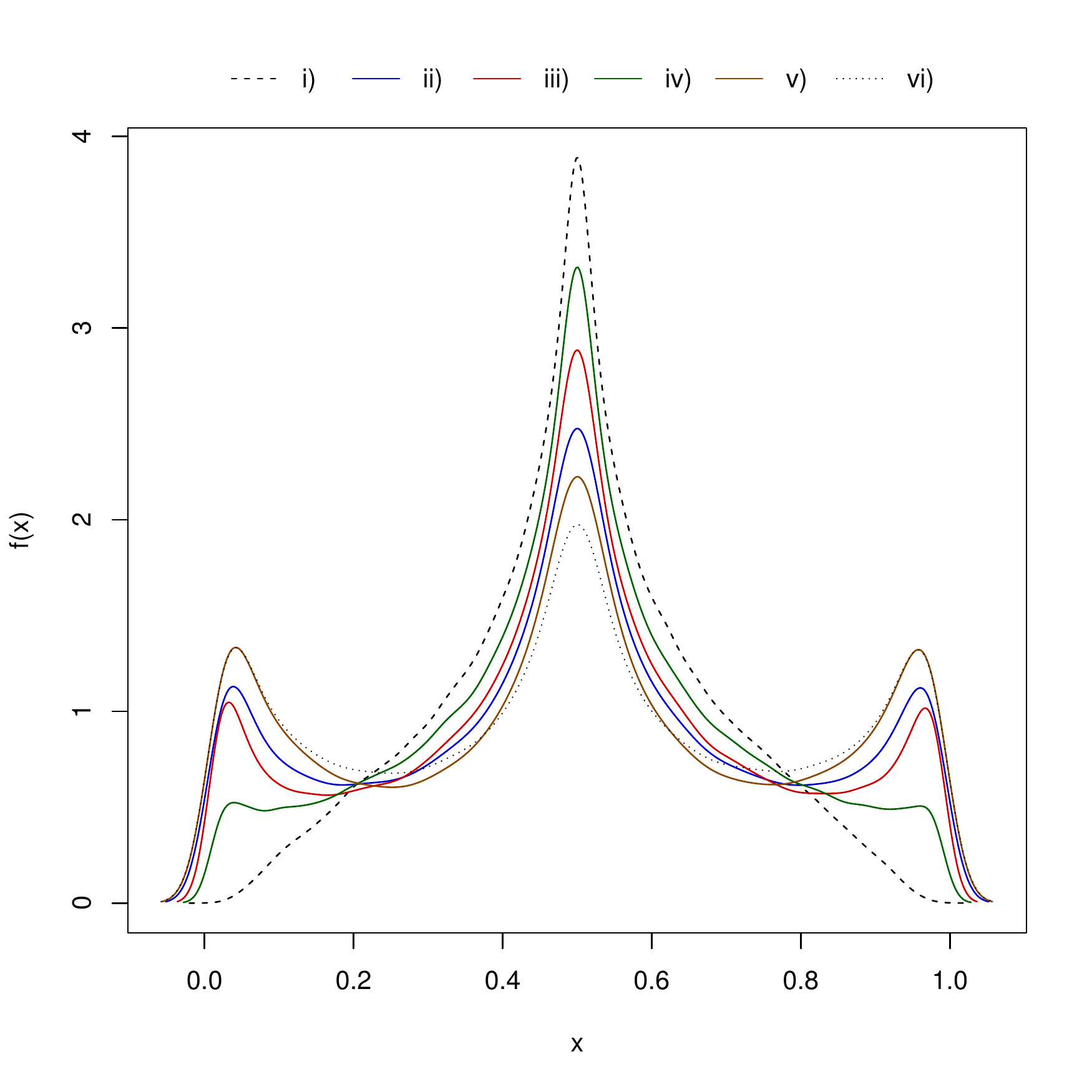}
	}\\
	\subfloat[$N=250$, $\tau=0.15$]{
		\includegraphics[scale=0.35]{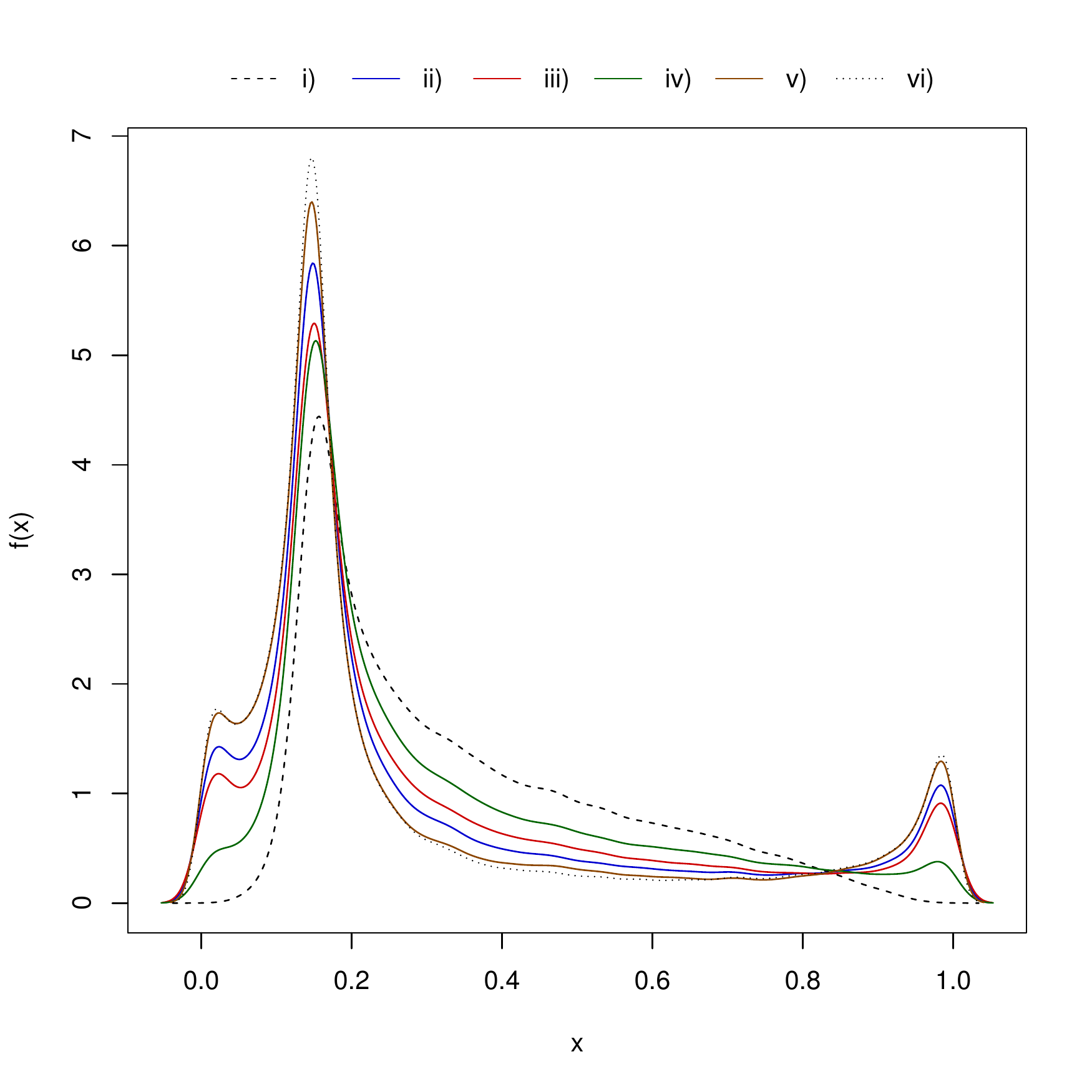}
	}
	\subfloat[$N=250$, $\tau=0.5$]{
		\includegraphics[scale=0.35]{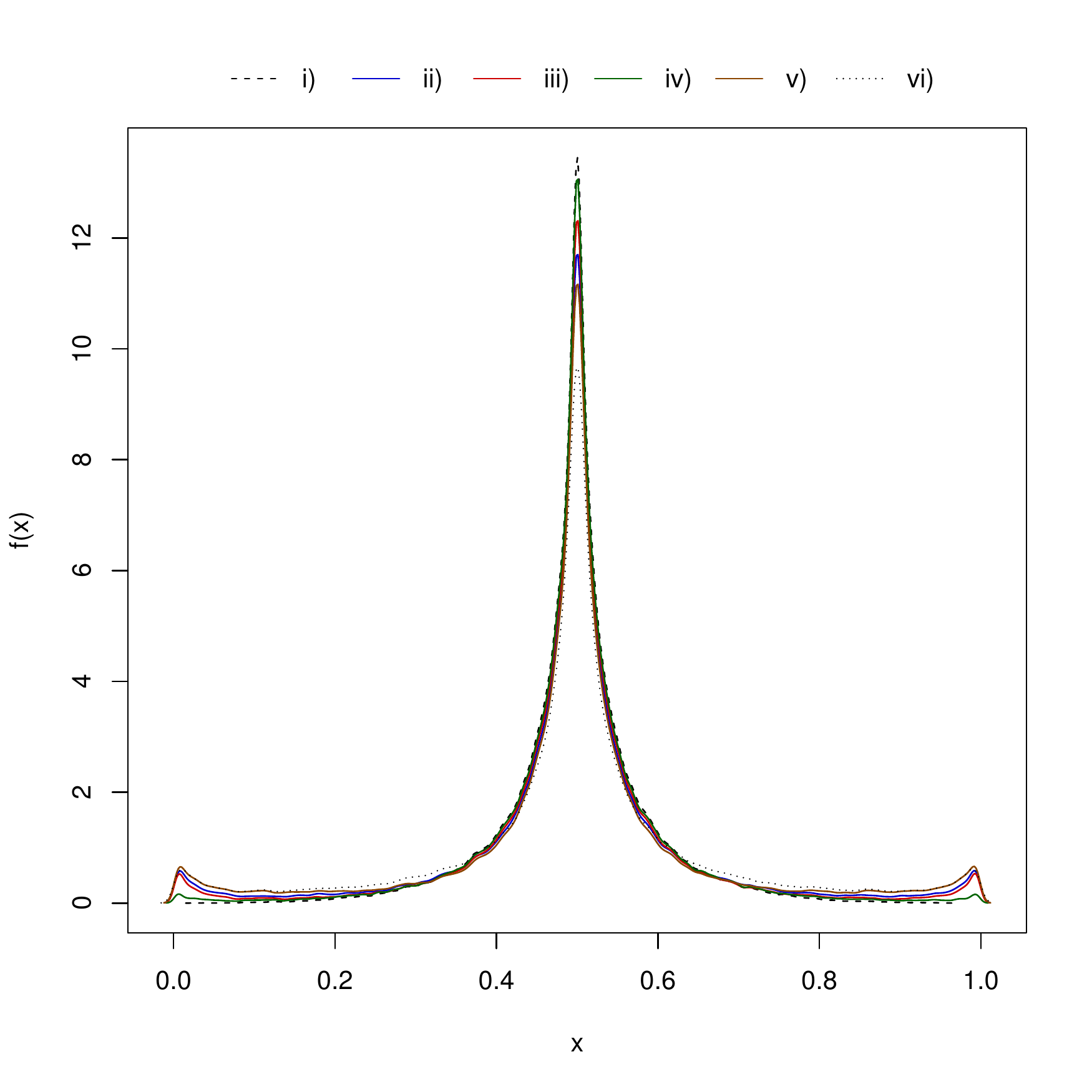}
	}
	\caption{empirical MSE ($M=10^5$ repetitions) of the plug-in change-point estimator for i.i. $U([0,1])$-distributed observations with change of size $d=0.7$ at $\tau=0.15$ (A) and (C) or $\tau=0.5$ (B) and (D) }
\end{figure}
\begin{figure}[H]
	\subfloat[$\tau=0.15$]{
		\resizebox{0.48\textwidth}{!}{\input{U_0_1_MSE_lambda=15pc.tex}}
	}
	\subfloat[$\tau=0.5$]{
		\resizebox{0.48\textwidth}{!}{\input{U_0_1_MSE_lambda=50pc.tex}}
	}
	\caption{empirical MSE ($M=10^5$ repetitions) of the plug-in change-point estimator for i.i. $U([0,1])$-distributed observations with change of size $d$ at $\tau=0.15$ (A) or $\tau=0.5$ (B) }
\end{figure}
\end{appendix}

\end{document}